\documentclass[10pt,twocolumn,letterpaper]{article}

\usepackage{cvpr}              % To produce the CAMERA-READY version
\usepackage{epsfig}
\usepackage{graphicx}
\usepackage{amsmath}
\usepackage{amssymb}

\usepackage[pagebackref=true,breaklinks=true,colorlinks,bookmarks=false]{hyperref}

\usepackage{amsfonts}
\usepackage{amsmath}
\usepackage{booktabs}
\usepackage{color}
\usepackage{url}
\usepackage{graphicx}
\usepackage{bbding}
\usepackage{multirow}
\usepackage{makecell}
\usepackage{newfloat}
\usepackage{listings}

\usepackage[marginal]{footmisc}

\usepackage[capitalize]{cleveref}
\crefname{section}{Sec.}{Secs.}
\Crefname{section}{Section}{Sections}
\Crefname{table}{Table}{Tables}
\crefname{table}{Tab.}{Tabs.}

\begin{document}

\title{DCANet: Differential Convolution Attention Network for RGB-D Semantic Segmentation}

% \author{Lizhi Bai \and
% Jun Yang \and
% Chunqi Tian\thanks{Corresponding author.}\and
% Yaoru Sun\thanks{Corresponding author.}\and
% Maoyu Mao\and
% Yanjun Xu\and
% Weirong Xu\\
% Department of Computer Science and Technology, Tongji University, Shanghai, 201804, China\\
% {\tt\small \{bailizhi, junyang, yaoru, maomy, jesse, 2132983\}@tongji.edu.cn, tianchunqi@163.com}
% }

% \author{Lizhi Bai^{$\dagger$} \and
% Jun Yang^{$\dagger$} \and
% Chunqi Tian^{$*$}\and
% Yaoru Sun^{$*$}\and
% Maoyu Mao\and
% Yanjun Xu\and
% Weirong Xu\\
% Department of Computer Science and Technology, Tongji University, Shanghai, 201804, China\\
% {\tt\small \{bailizhi, junyang, yaoru, maomy, jesse, 2132983\}@tongji.edu.cn, tianchunqi@163.com}
% }

\author{$\text{Lizhi Bai}^{\dagger}$ \and
$\text{Jun Yang}^{\dagger}$\and
$\text{Chunqi Tian}^{*}$\and
$\text{Yaoru Sun}^{*}$\and
Maoyu Mao\and
Yanjun Xu\and
Weirong Xu\\
Department of Computer Science and Technology, Tongji University, Shanghai, 201804, China\\
{\tt\small \{bailizhi, junyang, yaoru, maomy, jesse, 2132983\}@tongji.edu.cn, tianchunqi@163.com}
}

\maketitle
\footnote{
{$*$} Corresponding author. \\
{$\dagger$} Equal contribution.}

%%%%%%%%% ABSTRACT
\begin{abstract}
   Combining RGB images and the corresponding depth maps in semantic segmentation proves the effectiveness in the past
   few years. Existing RGB-D modal fusion methods either lack the non-linear feature fusion ability or treat both modal
   images equally, regardless of the intrinsic distribution gap or information loss. Here we find that depth maps are suitable to provide intrinsic fine-grained patterns of objects due to their local depth continuity, while RGB images
   effectively provide a global view. Based on this, we propose a pixel differential convolution attention (DCA) module
   to consider geometric information and local-range correlations for depth data. Furthermore, we extend DCA to ensemble
   differential convolution attention (EDCA) which propagates long-range contextual dependencies and seamlessly
   incorporates spatial distribution for RGB data. DCA and EDCA dynamically adjust convolutional weights by
   pixel difference to enable self-adaptive in local and long range, respectively. A two-branch network built with DCA
   and EDCA, called Differential Convolutional Network (DCANet), is proposed to fuse local and global information
   of two-modal data. Consequently, the individual advantage of RGB and depth data are emphasized. Our DCANet is shown
   to set a new state-of-the-art performance for RGB-D semantic segmentation on two challenging benchmark datasets,
   $i.e.,$ NYUDv2 and SUN-RGBD.
\end{abstract}

\begin{figure}
   \centering
   \includegraphics[scale=0.3]{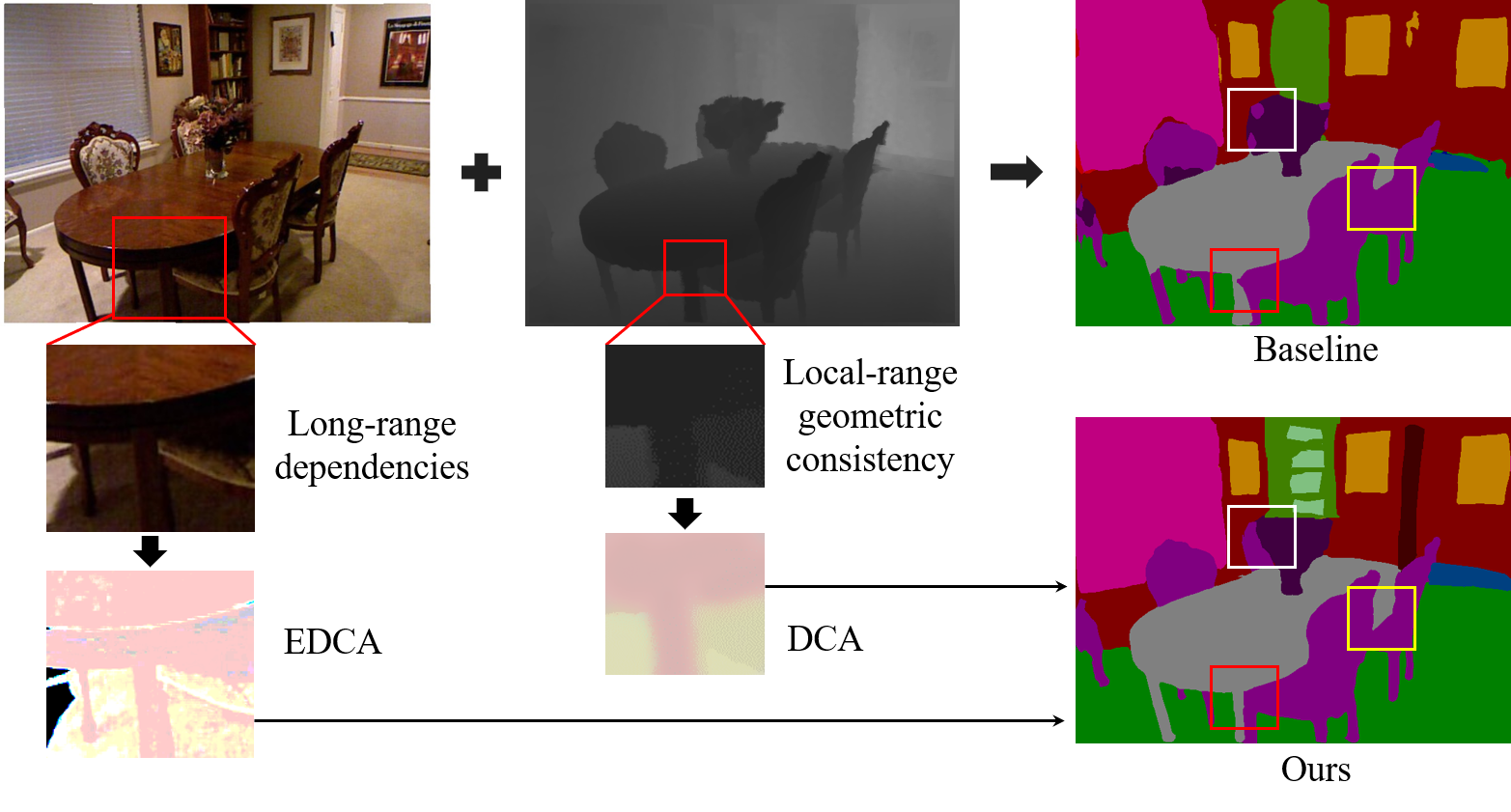}
   \caption{The intrinsic differences between RGB and depth data and the illumination of our DCANet.
      While the chair and the table are inseparable according to the 2D appearance in RGB image, they can be easily
      distinguished in depth map based on geometric information. In DCANet, we exploit DCA to
      capture local-range geometric consistency in depth map and EDCA to focus on long-range dependence for RGB.
   }
   \label{hard_sample2}
\end{figure}

%%%%%%%%% BODY TEXT
\section{Introduction}
Semantic segmentation is an essential task in computer vision, which infers semantic labels of every pixel in a scene.
With the widespread use of 3D sensors such as Kinect, Xition etc., the 3D geometry information of objects can be easily
obtained to boost the advancement of RGB-D semantic segmentation. After encoding the real-world geometric information,
the RGB-D images can be applied to overcome the challenge of 2D only displaying the photometric appearance properties
in the projected image space and enrich the representation of RGB images. The information of RGB and depth images are
presented in entirely different forms. In particular, RGB images capture the photometric appearance properties in the
projected image space, while the depth maps can produce plentiful complementary information for the appearance cues
of local geometry. As a result, it is vital to enhance and fuse the strengths of RGB and depth data in semantic
segmentation task.

In a real scenario, there are too many challenging images with complex appearances.
Take Fig.~\ref{hard_sample2} as an example, while
the chair and the table are inseparable in the RGB image, they can be easily distinguished in depth. Obviously, it is
not feasible to separate the table with chair using only 2D information such as shapes and colors.
While in the view of depth maps, there is local consistency information, which will not be limited by the similar confusing
appearance. In fact, the
depth data provide more fine-grained local geometry difference information and theoretically leading to better segmentation performance compared to only using RGB images. In contrast, as verified in the classic self-attention \cite{wang2018non,zhu2019asymmetric,zhang2019self} mechanisms that RGB data focuses on more global information.
% Nevertheless, CNN kernels are optimized by starting from random initialization, with no explicit encoding for gradient
% information, making them hard to focus on pixel difference.

The existing methods \cite{cao2021shapeconv,chen2020bi,chen2021spatial,mei2021depth,he2017std2p,husain2016combining,
   jiang2018rednet,cheng2017locality,lin2017cascaded, park2017rdfnet} try to fuse RGB-D data by introducing new convolution
layer and pooling layer, attention mechanism, noise-cancelling module, etc., to obtain better semantic segmentation results.
% are defined as RGB-D fusion methods, 
% the fusion method is too simple to take advantages
% of RGB images and depth maps. 
These methods ignore the intrinsic differences between RGB and depth features, using homogeneous operators instead.
The weights of both types of data are equally treated so as to make
the same contribution to the segmentation, which is obviously not appropriate. Besides, the information of RGB
images and depth maps is mainly achieved from the combined final channel, where specific semantic information in
different channels is not considered.

To address the aforementioned problems, we propose two attention mechanisms, namely differential convolution
attention (DCA) and ensemble differential convolution attention (EDCA) to improve the cross-modal ability between RGB and
depth data in semantic segmentation. DCA dynamically augments the standard convolution with a pixel difference term and forces pixels
with a similar difference to the center of the kernel to contribute more to the output than other pixels. DCA incorporates local
geometric information and improve local-range adaptability for depth data.
EDCA absorbs the advantage of dynamic convolution of DCA to propagate long-range contextual dependencies and seamlessly
incorporate spatial distribution for RGB data.
Meanwhile, both DCA and EDCA avoid common drawbacks such as ignoring adaptability in channel dimension. Our main contributions are summarized as follows.

$\bullet$ We propose a DCA module which incorporates local-range intricate geometric patterns and enables self-adaptive by considering
subtle discrepancy of pixels in local regions for depth data.

$\bullet$ We extend DCA to EDCA for achieving long-range correlations and seamlessly incorporating spatial distribution for RGB data.

$\bullet$ Based on DCA and EDCA, we propose a DCANet that achieves a new state-of-the-art performance on NYUDv2\cite{silberman2012indoor}
and SUN-RGBD\cite{song2015sun} datasets. We also provide a detailed analysis of design choices and model variants.

%------------------------------------------------------------------------

\section{Related Work}
\subsection{RGB-D Semantic Segmentation} With the help of additional depth information, the combination of such two complementary
modalities achieves great performance in semantic segmentation \cite{chen2021spatial,ren2012rgb,silberman2012indoor,jiao2019geometry,cao2021shapeconv,
   gupta2013perceptual,khan2016integrating}. Many works simply concatenate the features of RGB and depth images to enhance the semantic
information of each pixel \cite{silberman2012indoor,ren2012rgb}. The fusion method can be classified into three types: early fusion,
middle fusion and late fusion. Cao \textit{et~al.} \cite{cao2021shapeconv} concatenate the RGB and depth data decomposed by a shape
and a base component in the depth feature in the early stage. However, due to the complexity of these two modalities, a single model
cannot fit their data well due to their differences. Jiao \textit{\textit{et~al.}} \cite{jiao2019geometry} design two encoder-decoder
modules for fully consideration the RGB and depth information, where both modal are fused in late stage. In this method, the
interaction between the different features of RGB and depth data is insufficient, since the rich information of the modalities is
gradually compressed and even lost. After overcoming the drawback of early stage and late stage fusion strategy, middle stage fusion
performs better by fusing the intermediate information of the two different modalities. Gupta \textit{et~al.} \cite{gupta2014learning}
concatenate the geocentric embedding for depth images and with depth images to contribute the final semantic information in the
middle stage. Notably, the distribution gap is reduced in the middle stage fusion strategy, and multi-modal features are combined
with ample interaction. As a result, recent studies mainly focus on middle stage fusion. Chen \textit{et~al.} \cite{chen2021spatial}
propose a spatial information-guided convolution, which generates convolution kernels with different sampling distributions to enhance
the spatial adaptability of network and receptive field regulation. Chen \textit{et~al.} \cite{chen2020bi} unify the most informative
cross-modality features from data for both modalities into an efficient representation. Lin \textit{et~al.} \cite{lin2017cascaded}
split the image into multiple branches based on geometry information, where each branch of the network semantically segments relevant
similar features.

Our method applies two branches and each branch focuses on extracting modality-specific features, such as color and texture from RGB
images and geometric, illumination-independent features from depth images. To be specific, similar to middle stage fusion, attentive
depth features generated by the DCA are fused into the attentive RGB from the EDCA at each of the resolution stages in the
encoders. The depth and RGB data focus on local and long range information, respectively.

\subsection{Attention Modules} What has greatly contributed to the popularity of attention modules is the fact that they can be
applied to model the global dependencies of features almost in any stage of the network. Woo \textit{et~al.}
\cite{woo2018cbam} adaptively refined the information in spatial and channel dimensions through the convolutional block attention module.
Inspired by the self-attention network in Natural Language Processing \cite{vaswani2017attention}, such self-attention related module
achieves widespread focus in computer vision \cite{zhao2020exploring,vaswani2021scaling,ramachandran2019stand}. Many researchers focus
on the global and local dependencies.  In \cite{wang2018non}, Wang \textit{et~al.} propose a non-local model to extend the self-attention
to a more general type of non-local filtering method for capturing the long-range dependencies. Fu \textit{et~al.} \cite{fu2019dual}
propose two attention modules to capture spatial and channel interdependencies, respectively. Cao \textit{et~al.} \cite{cao2019gcnet}
propose a lightweight non-local network based on a query independent formulation for global context modeling.
Zhu \textit{et~al.} \cite{zhu2019asymmetric} integrates the features of different levels while considering long-range dependencies and
reducing redundant parameters.

Our method integrates DCA and EDCA to build relationship between different points for depth and RGB deta, respectively.
The DCA module supports that the same objects have more substantial depth similarity in a local-range of depth data, and we make use of the
pixel-wise difference to force pixels with more consistent geometry to make more contributions to the corresponding output.
The EDCA module enables long-range dependencies for RGB data.

\begin{figure}[t]
   \centering
   \includegraphics[scale=0.62]{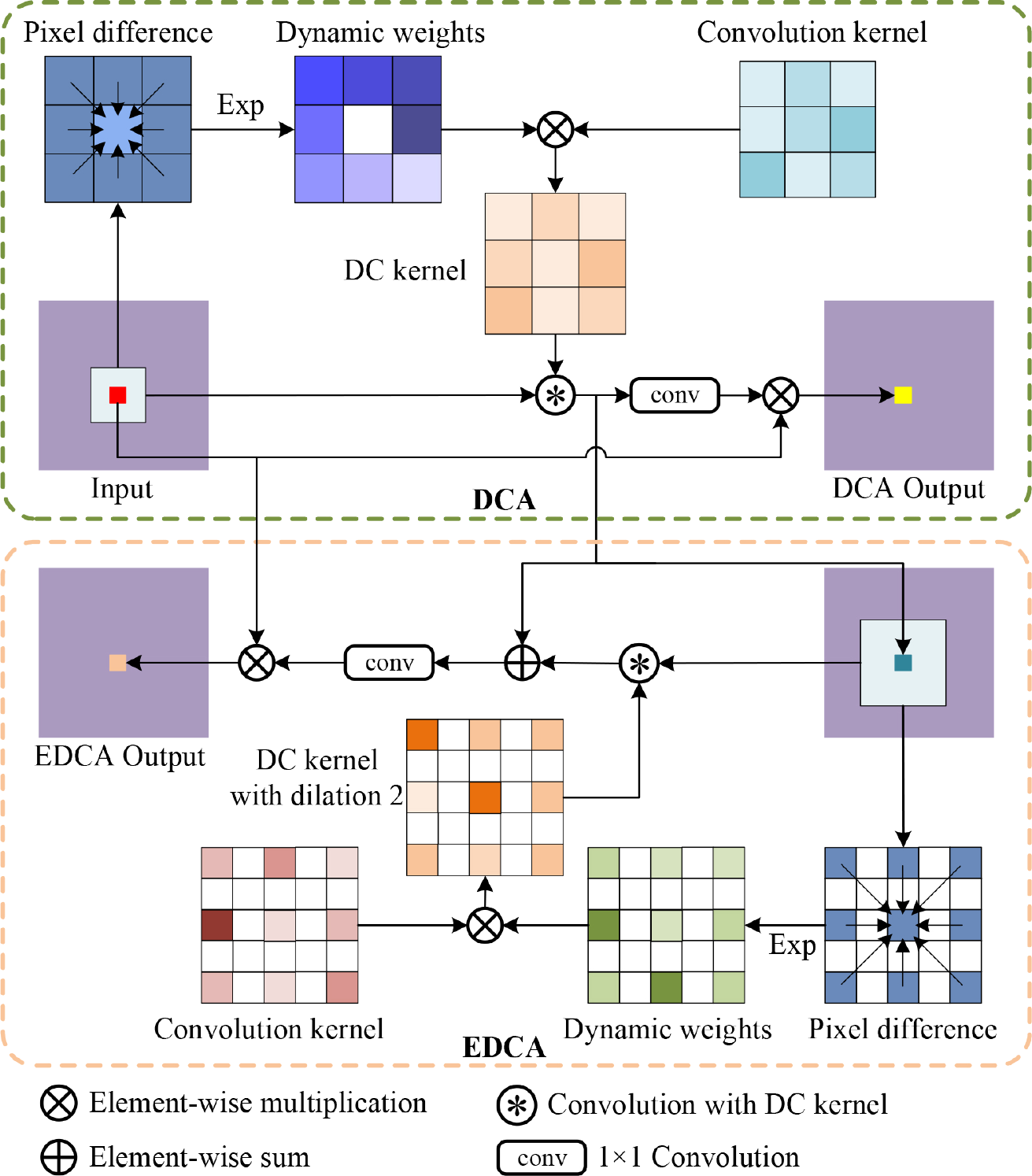}
   \caption{The instances of DCA and EDCA when taking a $3\times 3$ local grid as an example.}
   \label{piexl_diff}
\end{figure}

\section{Method}
RGB-D semantic segmentation requires fusing features from RGB and depth modalities, which are inherently different.
Specifically, RGB data has long-range contextual dependencies and global spatial consistency, while depth data contains local geometric
consistency. The intrinsic characteristics of the two modalities should be considered separately to identify the strengths of each, while
enhancing the two feature representations. To this end, we put forward two attention modules called DCA and EDCA to capture the intrinsic
features of depth and RGB data, respectively. In this section, we elaborate the details of the proposed DCA and EDCA, followed
by the the description of the proposed differential convolution attention network (DCANet).

\begin{figure}[t]
   \centering
   \includegraphics[scale=0.63]{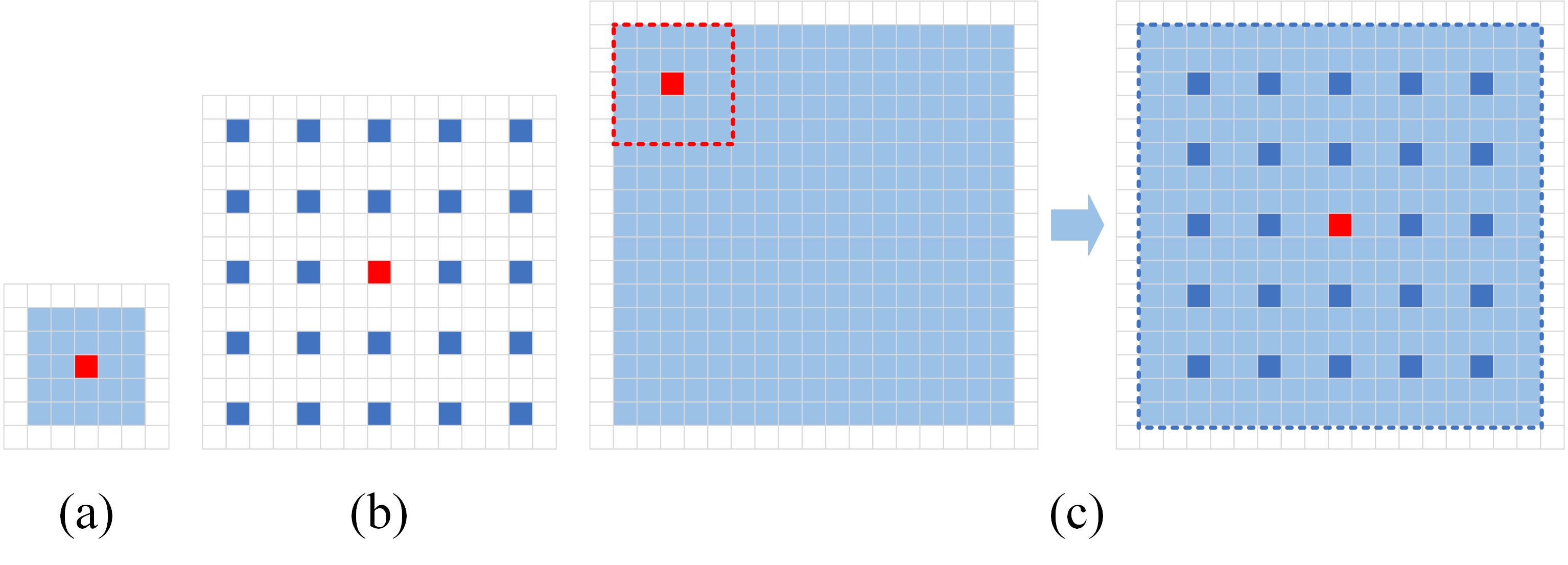}
   \caption{The explains of convolution strategies in EDC, $5\times 5$ convolution is used for convenience.
      (a) $5\times 5$ convolution, $\mathrm{Conv_{5\times 5}}$.
      (b) $5\times 5$ convolution with dilation 3, $\mathrm{DConv_{5\times 5}}$. (c) The combination of (a) and (b),
      $\mathrm{DConv_{5\times 5}(Conv_{5\times 5}( \cdot ))}$. Compared with (a),
      (b) has a larger receptive field, but cause information lost. In (c) (left), the red dashed
      box is $\mathrm{Conv_{5\times 5}}$ which just fills the dilation of $\mathrm{DConv_{5\times 5}}$ and makes
      it approximate size of $17\times 17$ as the blue dashed box in (c) (right). 
   }
   \label{dilatedconv}
\end{figure}

\subsection{Differential Convolution Attention}
\label{sec-DCA}
The attention mechanism can be considered as an adaptive selection process that selects discriminative features based on input features and
automatically ignores noisy responses \cite{guo2022visual}. The key point of the attention mechanism is to learn the relationship between
different points and generate an attention map that indicates the importance of different points. The well-known method for establishing
relationship between different points is self-attention mechanism \cite{wang2018non,xie2021segformer,fu2019dual,zhao2020exploring}, which
is used to capture long-range dependence. However, due to its intrinsic properties, the depth data is only relevant in a local region and
long-range dependencies may introduce more interference terms. For this, we explore convolution to build relevance and produce
attention map by considering a local region in depth data.

% It is essential to produce an attention map that shows the importance of different points in the attention mechanism. To do so, we should
% obtain the relationship between different points. In this paper, the pixel relationship (difference) is well calculated by our DDCA.

% \subsubsection{Vanilla Convolution.} 
Given a feature map $\mathbf{F} \in \mathbb{R}^{h\times w \times c}$; $h$, $w$, and $c$ are the height, width
and the channel of input feature map, respectively. For simplicity, we note $\mathbf{X} \in \mathbb{R}^{h\times w \times 1}$ as the input
feature map. For each point $\mathbf{p} \in \mathbb{R}^2$ on $\mathbf{X}$, the vanilla convolution is calculated as:
\begin{equation}
   \mathbf{Y}(\mathbf{p}) = \sum_{i=1}^{k\times k}\mathbf{K}_i\cdot \mathbf{X}(\mathbf{p}+\mathbf{p}_i),
   \label{eq1}
\end{equation}
where $\mathbf{p}_i$ enumerates the local locations around $\mathbf{p}$. $\mathbf{K}$ is the learnable weights of the convolution kernel with the size of $k \times k$ (the bias terms are ignored for
simplicity).

In Eq.(\ref*{eq1}), the convolution kernel $\mathbf{K}$ of the vanilla convolution is fixed for any input,
which cannot perceive the changes of the input dynamically. However, for depth data, we expect the attention map generated by convolution to
sense the geometric information on-the-fly while learning the correlations between different points in a local region. To this end,
we explore a pixel difference term to weight the vanilla convolution kernel called differential convolution kernel $\mathbf{K}^*$:
\begin{equation}
   \mathbf{K}^*_i = \mathbf{K}_i\cdot \mathrm{exp}(-|\mathbf{X}(\mathbf{p})-\mathbf{X}(\mathbf{p}+\mathbf{p}_i)|),
   \label{eq2}
\end{equation}
The difference term in $\mathbf{K}^*$ implies the geometric information in depth data
and is then regularized to (0,1], which ensures that the larger the difference between any two points, the smaller the correlation,
and vice versa. It is intuitive that the depth at a point is locally continuous.
With the blessing of the difference term, the differential convolution kernel $\mathbf{K}^*$ depends not only on the input features,
but also on the convolution position. Thus it is geometry-aware for depth data. With differential convolution kernel $\mathbf{K}^*$, the differential convolution (DC) for input feature map
$\mathbf{X} \in \mathbb{R}^{h\times w \times 1}$ can be written as:
\begin{equation}
   \mathbf{Y}(\mathbf{p}) = \sum_{i=1}^{k\times k}\mathbf{K}_i^*\cdot \mathbf{X}(\mathbf{p}+\mathbf{p}_i),\\
   \label{eq3}
\end{equation}
As mentioned above, we use differential convolution kernel $\mathbf{K}^*$ to calculate the relevance between different points
in a local receptive field, and the field size is input-dependent. In our experiments, the receptive field size for depth data
is $9 \times 9$. To reduce computations, we apply the depth-wise separable
convolution \cite{chollet2017xception} to decouple a differential convolution into a differential depth-wise convolution and a
point-wise convolution ($1\times 1$ convolution). For generalized input feature map $\mathbf{F} \in \mathbb{R}^{h\times w \times c}$,
our DCA module is defined as:
\begin{equation}
   \begin{split}
      Attention & =\mathrm{Conv_{1\times 1}(DC\mbox{-}DW(\mathbf{F}))}, \\
      Output    & =Attention\otimes \mathbf{F}.
   \end{split}
   \label{eq4}
\end{equation}
Here, $\mathrm{Conv_{1\times 1}}$ represents $1\times 1$ convolution, and $\mathrm{DC\mbox{-}DW}$ denotes differential depth-wise
convolution whose differential kernel is generated by Eq.(\ref*{eq2}).
$Attention\in \mathbb{R}^{h\times w \times c}$ means attention map which has the same size of input feature map $\mathbf{F}$.
Each value in the attention map integrates the geometric information in the local range of the depth image to indicate the
importance of each feature. $\otimes$ denotes element-wise product. The whole process of DCA is illustrated at the top part of Fig.~\ref{piexl_diff}.

Convolution kernel that introduces difference a term can dynamically rebalance the convolution weights according to the input.
And the proposed DCA module forces points with more consistent geometry to make more contributions to the corresponding output for depth data.
In summary, DCA achieves flexibility not only in the local spatial dimension, but also in the channel dimension, and integrates
geometric information of the local extent. It is worth noting that channels-wise information often represents different objects in CNNs
\cite{qin2020ffa,chen2017sca}, which is also crucial for segmentation tasks.

\subsection{Ensemble Differential Convolution Attention}
As mentioned above, RGB data has long-range contextual dependencies and global spatial consistency.
Although self-attention \cite{wang2018non,zhu2019asymmetric,zhang2019self} is the practical methods to learn relationship
between different points to capture long-range dependence, it only obtains spatial-wise adaptability
and lacks the channel-wise adaptability. The proposed
DCA module has flexibility in both spatial dimension and channel dimension, and it considers local-range
correlations which is appropriate for depth data. Therefore, as for RGB data it is intuitive to extend DCA to propagate long-range contextual
dependencies.

The most straightforward approach is to use larger kernel differential depth-wise convolution in DCA. In order to capture long-range relationship with less
computational costs and parameters than directly apply larger
kernel operations, we decomposed large kernel-based DC to a differential depth-wise convolution, a differential
depth-wise dilation convolution and a point-wise convolution, called ensemble differential convolution (EDC). With EDC, the
propose EDCA can be written as:
\begin{equation}
   \begin{split}
      \mathbf{F}_1&=\mathrm{DC\mbox{-}DW(\mathbf{F})},\\
      \mathbf{F}_2&=\mathrm{DC\mbox{-}DWD(\mathbf{F_1})},\\
      Attention & =\mathrm{Conv_{1\times 1}(\mathbf{F}_1+\mathbf{F}_2)}, \\
      Output    & =Attention\otimes \mathbf{F}.
   \end{split}
   \label{eq5}
\end{equation}
Similar to DCA, $\mathbf{F} \in \mathbb{R}^{h\times w \times c}$ is the input feature map. $\mathrm{Conv_{1\times 1}}$ represents
$1\times 1$ convolution and $\otimes$ denotes element-wise product. $\mathrm{DC\mbox{-}DW}$ and $\mathrm{DC\mbox{-}DWD}$
mean differential depth-wise convolution and differential depth-wise dilation convolution with differential convolution kernel
$\mathbf{K}^*$, respectively. Fig.~\ref{piexl_diff} shows the proposed EDCA module.

The size of EDC kernels are also input dependent. In our experiments, the DC kernel size of
$\mathrm{DC\mbox{-}DW}$ is $5\times 5$, and that of $\mathrm{DC\mbox{-}DWD}$ is $9\times 9$ with dilation 3. With the above
settings, the receptive field size of EDC is approximated to $29\times 29$.
Fig.~\ref{dilatedconv} (d) shows the convolution strategies in EDC, for convenience, we show the $5\times 5$ convolution and $5\times 5$ convolution
with dilation 3. Accordingly,
EDCA can obtain long-range dependence, while the differential term dynamically adjust the convolution weights and
provides spatial distribution information for RGB data. In summary, the discriminative features are boosted and the noisy
responses are ignored based on the spatial and channel-wise adaptability of EDCA.

\subsection{Understanding DCA and EDCA}
As verified by the prior works that pixels with same semantic labels have similar depths \cite{lin2017cascaded, mei2021depth,
   wang2018depth} in a local region. The DCA integrates geometric perception ability to vanilla convolution and generates an attention map
which indicates the importance of each point in depth data. EDCA absorbs the advantage of dynamic convolution of DCA to propagate
long-range contextual dependencies and seamlessly incorporate spatial distribution for RGB data.

\begin{table}[]
   \centering
   \caption{Desirable characteristics of convolution, self-attention, DCA, and EDCA. Notably, DCA and EDCA are applied for depth and RGB data, respectively.}
   \resizebox{1\linewidth}{!}{
      \begin{tabular}{c|c|c|c|c}
         \hline
         \multicolumn{1}{c|}{Properties}            & Convolution                         & \multicolumn{1}{c|}{self-attention} & DCA                               & EDCA           \\ \hline
         Geometry Structure                         & \multicolumn{1}{c|}{\XSolidBrush}   & \XSolidBrush                        & \CheckmarkBold                    & \XSolidBrush   \\
         Local-range dependence                     & \multicolumn{1}{c|}{\CheckmarkBold} & \XSolidBrush                        & \CheckmarkBold                    & \CheckmarkBold \\
         \multicolumn{1}{c|}{Long-range dependence} & \XSolidBrush                        & \multicolumn{1}{c|}{\CheckmarkBold} & \multicolumn{1}{c|}{\XSolidBrush} & \CheckmarkBold \\
         Spatial adaptability                       & \XSolidBrush                        & \multicolumn{1}{c|}{\CheckmarkBold} & \CheckmarkBold                    & \CheckmarkBold \\
         Channel adaptability                       & \XSolidBrush                        & \XSolidBrush                        & \CheckmarkBold                    & \CheckmarkBold \\ \hline
      \end{tabular}}
   \label{prop}
\end{table}

As shown in Tab.~\ref{prop}, our proposed DCA and EDCA combine the advantages of convolution and self-attention. By augmenting the convolution kernel with a pixel difference
term, DCA captures geometry with a local receptive field. Compared with vanilla convolution, the learnable weights of DCA are adjusted
by the geometric variance.
Based on this, with the help of our decomposed large kernel, the EDCA is extended to further capture refined pixel discrepancy in the satisfied receptive field.

\begin{figure*}
   \centering
   % \begin{minipage}[c]{0.4\linewidth}
   \centering
   \centerline{
      \includegraphics[scale=0.45]{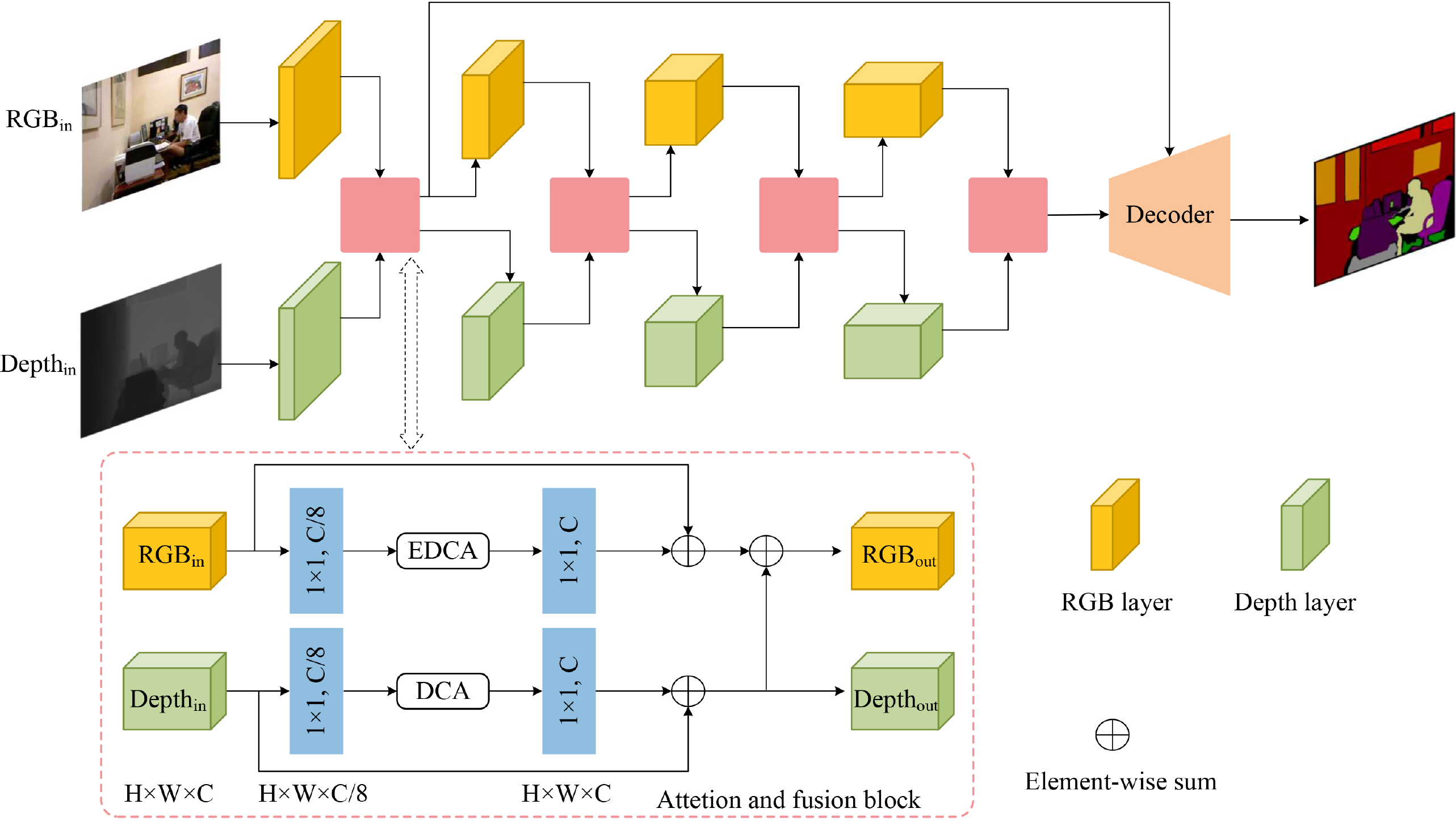}
   }
   \caption{The overview of our network. The network consists of two ResNet-101 encoders, where DCA and EDCA are plugged
      into CNNs as an attention module for each block of each ResNet-101 encoder in RGB and Depth branches, respectively. We employ the
      original decoder DeepLabv3+. During training, each pair of feature maps are fused by the attention and fusion block and propagated to the next stage of the
      encoder for further feature transformation.}
   \label{model_architecture}
\end{figure*}

% \section{Model Architecture}
\subsection{DCANet Architecture}
The architecture of DCANet for RGB-D semantic segmentation is shown in Fig.~\ref{model_architecture}. Our DCANet adopts
DeepLabv3+ \cite{chen2018encoder} as the baseline for RGB-D semantic segmentation task, where the encoder is
ResNet-101 \cite{he2016deep} and retain the original decoder of DeepLabv3+. We apply a two-branch structure in our DCANet,
one for RGB and another for depth data.
% and fuse the feature representations in our network.

% \subsection{Cross-modal fusion}
At each of the four resolution stages in the ResNet-101, depth features are fused into the RGB encoder by attention and fusion block. Specifically, the channel
dimension of both modalities are first squeezed to $1/8$ for dimensionality reduction. Next, we apply the DCA for depth data and EDCA for
RGB data simultaneously. Third, the outputs of DCA and EDCA are convolved to match the dimensionality of the original features
and performed an element-wise sum with the original features separately. Finally, the depth data extracting complementary geometric
information is integrated into the RGB data by element-wise sum to obtain better feature representations. The outputs of attention and
fusion block are as follows:

% As shown in Fig.~\ref{model_architecture}, the EDCA and DCA are employed in each attention and fusion block, where the output of RGB and depth data as follows:

\begin{equation}
   \begin{split}
      \text{Depth}_\text{out} &= \mathbb{W}_2\left(\text{DCA}(\mathbb{W}_1 (\text{Depth}_\text{in})) \right) + \text{Depth}_\text{in},\\
      \text{RGB}_\text{out} &= \mathbb{W}_2'\left( \text{EDCA} (\mathbb{W}_1' (\text{RGB}_\text{in})) \right) + \text{RGB}_\text{in},\\
      \text{RGB}_\text{out} &=\text{RGB}_\text{out} + \text{Depth}_\text{out}
   \end{split}
   \label{contribution}
\end{equation}
where $\mathbb{W}_1$ ($\mathbb{W}_1'$) and $\mathbb{W}_2$ ($\mathbb{W}_2'$) represent $1\times 1$ convolution to squeeze and recover the channel, respectively.
Notably, the fused output RGB feature of last block is propagated to the segmentation decoder.
% where $\mathbb{W}_1$ represents the Conv$_{1\times 1 \times C/8}$ to squeeze the channel and $\mathbb{W}_2$ represents Conv$_{1\times 1 \times C}$
% to recover the channel. Notably, the fused output RGB feature of last block is propagated to the segmentation decoder.

%-------------------------------------------------------------------------

\section{Experiments}
\label{Experiment}

\subsection{Dataset and metrics}
Evaluation is performed on two popular RGB-D datasets:

NYUDv2 \cite{silberman2012indoor}: NYUDv2 contains 1449 RGB-D images with pixel-wise labels.
We follow the 40-class settings and the standard split with 795 training images and 654 testing images.

SUN-RGBD \cite{song2015sun}: This dataset has 37 categories of objects and consists 10335 RGB-D images, with 5285 as training
and 5050 as testing.

We evaluate the results using two common metrics, $i.e.,$ Pixel Accuracy (Pixel Acc.), and Mean Intersection Over Union (mIoU).

\subsection{Implementation Details}
We use dilated ResNet-101\cite{he2016deep} pretrained on ImageNet\cite{russakovsky2015imagenet} as the backbone network for
feature extraction and adding another auxiliary loss in the last stage of ResNet-101. We keep all the other settings of
DeepLabv3+ \cite{chen2018encoder} the same.
We implement our network using the PyTorch deep learning framework \cite{paszke2019pytorch}, and all the models are trained
with two Nvidia Tesla V100 GPUs. We employ the ``poly” policy \cite{liu2015parsenet} with initial learning rate 0.008, crop
size $480 \times 480$, batch size $8$, fine-tuning batch normalization parameters \cite{ioffe2015batch} and data augmentation
method ($i.e.,$ random scaling, random cropping, and left-right flipping) during training. For the optimizer, we use the
SGD with a momentum of 0.9 and a weight decay of 0.0001. In addition, we train the NYUDv2 dataset for 500 epochs and train
the SUN-RGBD dataset for 200 epochs. For fair comparisons with other methods, we adopted both single-scale and multi-scale
testing strategies during inference. If not otherwise noted, the experiments are single-scale testing, and `$*$' in tables
denote the multi-scale strategy.

\subsection{Ablation Study}
\textbf{DC kernel size of DCA.}
% Our DCA applies the DC kernel of $9 \times9$, dilation 1 to capture local geometry information for depth data, while
% EDCA achieves the receptive field of $27 \times27$ to propagate long-range contextual dependencies and spatial distribution for
% RGB data. To confirm the effectiveness of applying $9 \times9$ DC kernel, we henceforth attempt other DC
% kernel size. The results shown in Tab.\ref{kernel_size} prove that our setup works.
Our DCA module applies the DC kernel of $9 \times9$, dilation 1 to capture local geometry information for depth data.
To confirm the effectiveness of applying $9 \times9$ DC kernel, we hence attempt DCA with other DC kernel
sizes on depth data and no operations are performed on the RGB data. The results shown in Tab.\ref{kernel_size}
prove that larger DC kernels do not bring significant performance gains due to
the local geometric nature of depth data and our setup works.

\begin{table}
   \centering
   \caption{The results of DCA with different DC kernel sizes on NYUDv2 test set.}
   \label{nyud2}
   \resizebox{0.5\linewidth}{!}{
      \begin{tabular}{c|cc}
         \toprule
         DC kernel size & Pixel Acc. & mIoU \\  \midrule
         $3 \times3$    & 75.3       & 49.1 \\
         $5 \times5$    & 75.7       & 49.7 \\
         $7 \times7$    & 76.0       & 50.1 \\
         $9 \times9$    & 76.5       & 50.9 \\
         $11 \times11$  & 76.4       & 50.9 \\  \bottomrule
      \end{tabular}}
   \label{kernel_size}
\end{table}

\begin{table}[]
   \centering
   \caption{Ablation study on DCA and EDCA modules on NYUDv2 test set.}
   \resizebox{0.75\linewidth}{!}{
      \begin{tabular}{c|cc|cc}
         \toprule
         Method   & DCA        & EDCA       & Pixel Acc.\%  & mIoU\%        \\ \midrule
         Baseline &            &            & 75.1          & 47.4          \\
         Model1   & \Checkmark &            & 76.5          & 50.9          \\
         Model2   &            & \Checkmark & 76.9          & 51.3          \\
         DCANet   & \Checkmark & \Checkmark & \textbf{77.3} & \textbf{52.1} \\  \bottomrule
      \end{tabular}}
   \label{Ablation1}
\end{table}

\textbf{Effectiveness of DCA and EDCA modules.}
We conduct ablation studies on NYUDv2 dataset to prove the indispensability of the DCA and EDCA modules. We perform two parallel DeepLabv3+ (ResNet-101) as baseline. As shown in Tab.~\ref{Ablation1},
the two attention modules improve the performance remarkably. Compared with baseline, employing only DCA on depth data
improves mIoU by 3.5\%, while using only EDCA on RGB data brings 3.9\% improvement. When we apply both modules together,
the performance is further improved to 77.3\% (Pixel Acc.) and 52.1\% (mIoU). The results indicate that both modules are
critical for the performance improvement and work best when combined.

\begin{table}[]
   \centering
   \caption{Superiority of EDCA compared with Self-Attention \cite{wang2018non} and EDCA- on NYUDv2 test set. EDCA- denotes
      EDCA without differential term. All the three modules are for RGB data to capture long-range dependence and no
      operations are performed on the depth data.}
   \resizebox{0.9\linewidth}{!}{
      \begin{tabular}{ccc|cc}
         \toprule
         Self-Attention & EDCA-      & EDCA       & Pixel Acc.\% & mIoU\% \\ \midrule
         \Checkmark     &            &            & 76.1         & 49.3   \\
                        & \Checkmark &            & 76.3         & 50.1   \\
                        &            & \Checkmark & 76.9         & 51.3   \\ \bottomrule
      \end{tabular}}
   \label{Ablation3}
\end{table}

\textbf{EDCA vs. Self-Attention.}
Self-attention mechanism, $e.g.$, Non-local neural networks \cite{wang2018non}, are the well-known method to capture
long-range dependence. We compare the performance of self-attention with our proposed EDCA. As illustrated in Tab.~\ref{Ablation3}, EDCA outperforms self-attention in mIoU and Pixel Acc.
by 2\%  and 0.8\%, respectively. Self-attentive mechanisms are spatially adaptive, but not simultaneously channel-adaptive as EDCA.
Nevertheless, channel adaptability plays a crucial role in segmentation tasks. Furthermore, we also verify the effectiveness
of differential term in EDCA by removing the differential term in EDCA, called EDCA-. The results in Tab.~\ref{Ablation3} show
that differential term brings 1.2\% improvement in mIoU. This term in EDCA provides long range spatial distribution
information for RGB data while dynamically sensing the scene.

\begin{table}[]
   \centering
   \caption{Suitability of DCA and EDCA of the NYUDv2 test set. Note: Our proposed method applies DCA for Depth and EDCA for RGB data.}
   \resizebox{0.6\linewidth}{!}{
      \begin{tabular}{cc|cc}
         \toprule
         RGB  & Depth & Pixel Acc. & mIoU \\ \midrule
         EDCA &       & 76.9       & 51.3 \\
         DCA  &       & 76.2       & 49.7 \\
              & DCA   & 76.5       & 50.9 \\
              & EDCA  & 76.1       & 49.2 \\ \bottomrule
      \end{tabular}}
   \label{Ablation2}
\end{table}

\textbf{Suitability of DCA and EDCA.}
In the proposed DCANet, we apply DCA on depth data to capture local-range dependence and geometric information and EDCA on
RGB data to garner long-range correlations and spatial distribution information.
We also confirm the suitability of such two modules by applying EDCA for depth data and DCA for RGB. As shown in Tab.~\ref{Ablation2},
applying DCA on depth improves mIoU by 1.7\% compared to EDCA and 1.6\% improvement in mIoU using EDCA over DCA on RGB.
The results illustrate that DCA and EDCA are appropriate for Depth and RGB data, respectively. This also explains that depth
maps are more suitable for providing intrinsic geometric information of objects due to their local depth continuity, while RGB
images effectively provide a global view.

\begin{table}
   \centering
   \caption{Single-scale testing performance comparison with different baseline methods on NYUDv2 test set. }
   \resizebox{0.9\linewidth}{!}{
      \begin{tabular}{c|c|ccc}
         \hline
         Method                                                         & Backbone                    & Setting  & Pixel Acc. & mIoU \\ \hline
         \multirow{6}{*}{\makecell{Deeplabv3+\cite{chen2018encoder}}}   & \multirow{3}{*}{ResNet-101} & baseline & 75.1       & 47.4 \\ \cline{3-5}
                                                                        &                             & ours     & 77.3       & 52.1 \\ \cline{3-5}
                                                                        &                             & +        & 2.2        & 4.7  \\ \cline{2-5}
                                                                        & \multirow{3}{*}{ResNet-50}  & baseline & 74.5       & 46.5 \\ \cline{3-5}
                                                                        &                             & ours     & 76.8       & 51.2 \\ \cline{3-5}
                                                                        &                             & +        & 2.3        & 4.7  \\ \hline
         \multirow{6}{*}{\makecell{Deeplabv3\cite{chen2017rethinking}}} & \multirow{3}{*}{ResNet-101} & baseline & 73.3       & 45.4 \\ \cline{3-5}
                                                                        &                             & ours     & 76.2       & 50.2 \\ \cline{3-5}
                                                                        &                             & +        & 2.9        & 4.8  \\ \cline{2-5}
                                                                        & \multirow{3}{*}{ResNet-50}  & baseline & 72.7       & 45.2 \\ \cline{3-5}
                                                                        &                             & ours     & 75.8       & 49.4 \\ \cline{3-5}
                                                                        &                             & +        & 3.1        & 4.2  \\ \hline
         \multirow{6}{*}{\makecell{PSPNet\cite{lin2017feature}}}        & \multirow{3}{*}{ResNet-101} & baseline & 72.8       & 44.3 \\ \cline{3-5}
                                                                        &                             & ours     & 75.6       & 49.2 \\ \cline{3-5}
                                                                        &                             & +        & 2.8        & 4.9  \\ \cline{2-5}
                                                                        & \multirow{3}{*}{ResNet-50}  & baseline & 72.2       & 43.6 \\ \cline{3-5}
                                                                        &                             & ours     & 75.1       & 48.5 \\ \cline{3-5}
                                                                        &                             & +        & 2.9        & 4.9  \\ \hline
         \multirow{6}{*}{\makecell{FPN\cite{zhao2017pyramid}}}          & \multirow{3}{*}{ResNet-101} & baseline & 74.4       & 46.5 \\ \cline{3-5}
                                                                        &                             & ours     & 76.1       & 50.1 \\ \cline{3-5}
                                                                        &                             & +        & 1.7        & 3.6  \\ \cline{2-5}
                                                                        & \multirow{3}{*}{ResNet-50}  & baseline & 74.1       & 46.2 \\ \cline{3-5}
                                                                        &                             & ours     & 75.7       & 49.9 \\ \cline{3-5}
                                                                        &                             & +        & 1.6        & 3.7  \\ \hline
      \end{tabular}}
   \label{architectures}
\end{table}

\subsection{Experiments on Different Architectures}
Our proposed DCA and EDCA are general modules for RGB-D semantic segmentation, which can be easily plugged into CNNs
as attention modules in semantic segmentation. Our method is also assessed with respect to several
representative semantic segmentation architectures: Deeplabv3+ \cite{chen2018encoder}, Deeplabv3 \cite{chen2017rethinking},
PSPNet \cite{lin2017feature} and FPN \cite{zhao2017pyramid} with different backbones (ResNet-50, ResNet-101 \cite{he2016deep})
on NYUDv2 dataset to verify the generalizability. As is shown in Tab.~\ref{architectures}, our method outperforms the baseline
by a desirable margin under all settings, demonstrating the generalization capability of our method.

\begin{table}
   \centering
   \caption{Performance comparison with the state-of-the-art methods on NYUDv2 test set. `*' means multi-scale testing.}
   \resizebox{0.8\linewidth}{!}{
      \begin{tabular}{c|cc}
         \toprule
         Method                                   & Pixel Acc.(\%) & mIoU(\%)      \\ \midrule
         % FCN \cite{long2015fully}                 & 65.4           & 34.0          \\
         LSD-GF \cite{cheng2017locality}          & 71.9           & 45.9          \\
         D-CNN \cite{wang2018depth}               & -              & 48.4          \\
         MMAF-Net \cite{fooladgar2019multi}       & 72.2           & 44.8          \\
         ACNet \cite{hu2019acnet}                 & -              & 48.3          \\
         ShapeConv \cite{cao2021shapeconv}        & 75.8           & 50.2          \\
         RDF \cite{park2017rdfnet}*               & 76.0           & 50.1          \\
         M2.5D \cite{xing2020malleable}*          & 76.9           & 50.9          \\
         SGNet \cite{chen2021spatial}*            & 76.8           & 51.1          \\
         SA-Gate \cite{chen2020bi}*               & 77.9           & 52.4          \\
         InverseForm \cite{borse2021inverseform}* & 78.1           & 53.1          \\
         ShapeConv \cite{cao2021shapeconv}*       & 76.4           & 51.3          \\ \hline
         DCANet                                   & 77.3           & 52.1          \\
         DCANet*                                  & \textbf{78.2}  & \textbf{53.3} \\  \bottomrule
      \end{tabular}}
   \label{nyuv2}
\end{table}

\begin{table}
   \centering
   \caption{Performance comparison with the state-of-the-art methods on SUN RGB-D test set. `*' means multi-scale testing.}
   \resizebox{0.8\linewidth}{!}{
      \begin{tabular}{c|cc}
         \toprule
         Method                             & Pixel Acc.(\%) & mIoU(\%)      \\ \midrule
         3DGNN \cite{qi20173d}              & -              & 44.1          \\
         D-CNN \cite{wang2018depth}         & -              & 42.0          \\
         MMAF-Net \cite{fooladgar2019multi} & 81.0           & 47.0          \\
         SGNet \cite{chen2021spatial}       & 81.0           & 47.5          \\
         ShapeConv \cite{cao2021shapeconv}  & 82.0           & 47.6          \\
         ACNet \cite{hu2019acnet}           & -              & 48.1          \\
         3DGNN \cite{qi20173d}*             & -              & 45.9          \\
         CRF \cite{lin2017cascaded}*        & -              & 48.1          \\
         RDF \cite{park2017rdfnet}*         & 81.5           & 47.7          \\
         SA-Gate \cite{chen2020bi}*         & 82.5           & 49.4          \\
         SGNet \cite{chen2021spatial}*      & 82.0           & 47.6          \\
         ShapeConv \cite{cao2021shapeconv}* & 82.2           & 48.6          \\\hline
         % RefineNet \cite{lin2017refinenet}* & -              & 48.1          \\ \hline
         DCANet                             & 82.2           & 48.1          \\
         DCANet*                            & \textbf{82.6}  & \textbf{49.6} \\ \bottomrule
      \end{tabular}}
   \label{sunrgbd}
\end{table}

% 5+5 并排，各两列数据集
\begin{figure*}[t]
   \centering
   %1
%   \begin{minipage}[c]{0.091\linewidth}
%       \includegraphics[width=0.66in]{pic/visualize/nyuv2/395_rgb.jpg}
%   \end{minipage}
%   \begin{minipage}[c]{0.091\linewidth}
%       \includegraphics[width=0.66in]{pic/visualize/nyuv2/395_depth.png}
%   \end{minipage}
%   \begin{minipage}[c]{0.091\linewidth}
%       \includegraphics[width=0.66in]{pic/visualize/nyuv2/395_gt.png}
%   \end{minipage}
%   \begin{minipage}[c]{0.091\linewidth}
%       \includegraphics[width=0.66in]{pic/visualize/nyuv2/395_bs.png}
%   \end{minipage}
%   \begin{minipage}[c]{0.091\linewidth}
%       \includegraphics[width=0.66in]{pic/visualize/nyuv2/395_pred.png}
%   \end{minipage}\hspace{3pt}
%   \begin{minipage}[c]{0.091\linewidth}
%       \includegraphics[width=0.66in]{pic/visualize/sunrgbd/0000021_rgb.jpg}
%   \end{minipage}
%   \begin{minipage}[c]{0.091\linewidth}
%       \includegraphics[width=0.66in]{pic/visualize/sunrgbd/0000021_depth.png}
%   \end{minipage}
%   \begin{minipage}[c]{0.091\linewidth}
%       \includegraphics[width=0.66in]{pic/visualize/sunrgbd/0000021_gt.png}
%   \end{minipage}
%   \begin{minipage}[c]{0.091\linewidth}
%       \includegraphics[width=0.66in]{pic/visualize/sunrgbd/0000021_bs.png}
%   \end{minipage}
%   \begin{minipage}[c]{0.091\linewidth}
%       \includegraphics[width=0.66in]{pic/visualize/sunrgbd/0000021_pred.png}
%   \end{minipage}

   %2
   \begin{minipage}[c]{0.091\linewidth}
      \includegraphics[width=0.66in]{pic/visualize/nyuv2/350_rgb.jpg}
   \end{minipage}
   \begin{minipage}[c]{0.091\linewidth}
      \includegraphics[width=0.66in]{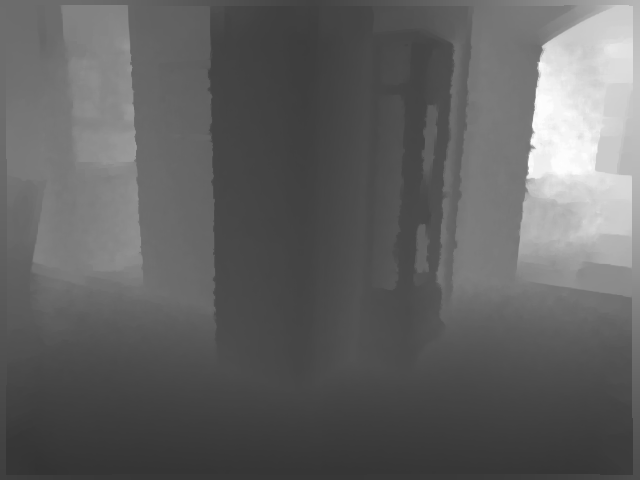}
   \end{minipage}
   \begin{minipage}[c]{0.091\linewidth}
      \includegraphics[width=0.66in]{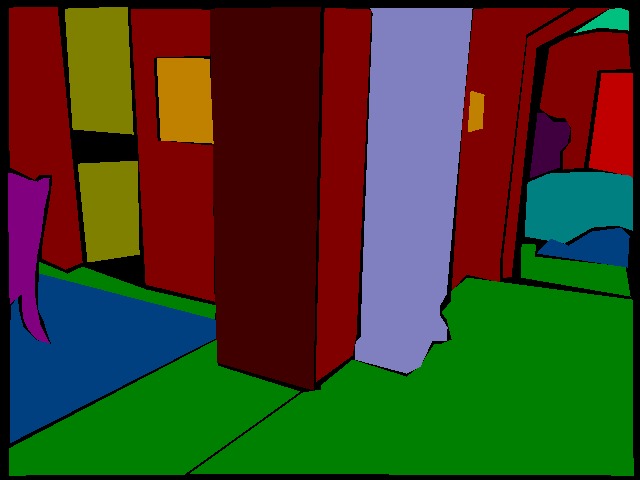}
   \end{minipage}
   \begin{minipage}[c]{0.091\linewidth}
      \includegraphics[width=0.66in]{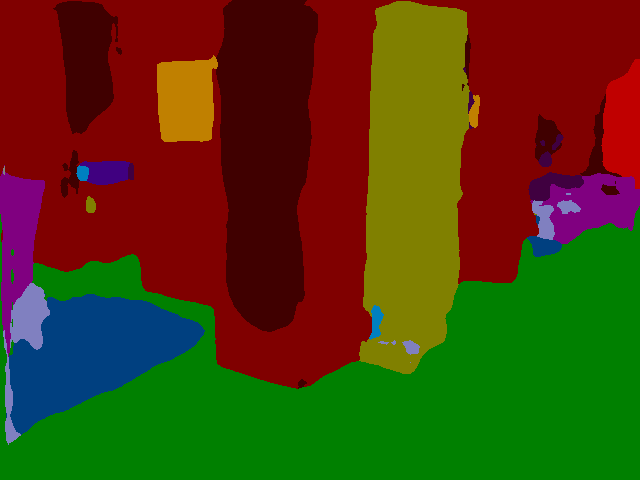}
   \end{minipage}
   \begin{minipage}[c]{0.091\linewidth}
      \includegraphics[width=0.66in]{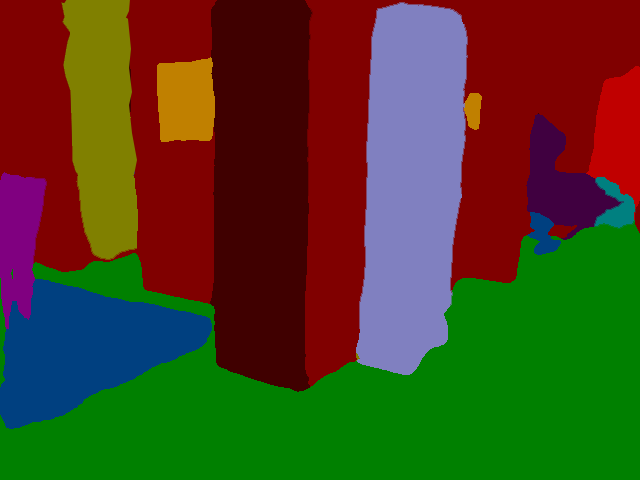}
   \end{minipage}\hspace{3pt}
   \begin{minipage}[c]{0.091\linewidth}
      \includegraphics[width=0.66in]{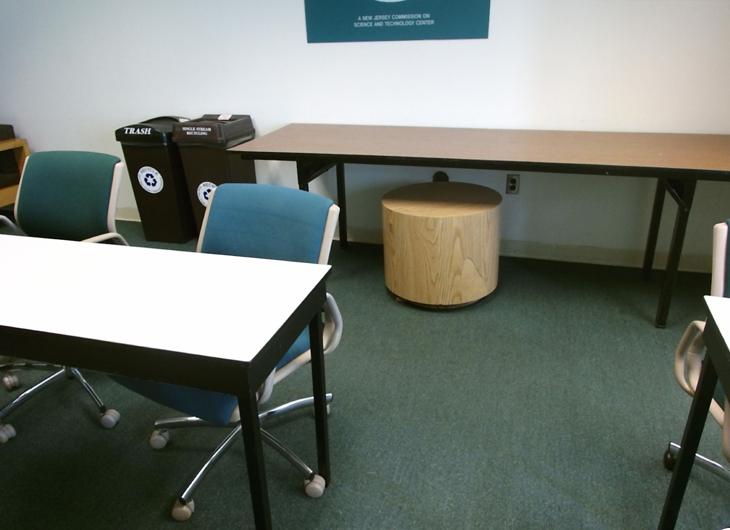}
   \end{minipage}
   \begin{minipage}[c]{0.091\linewidth}
      \includegraphics[width=0.66in]{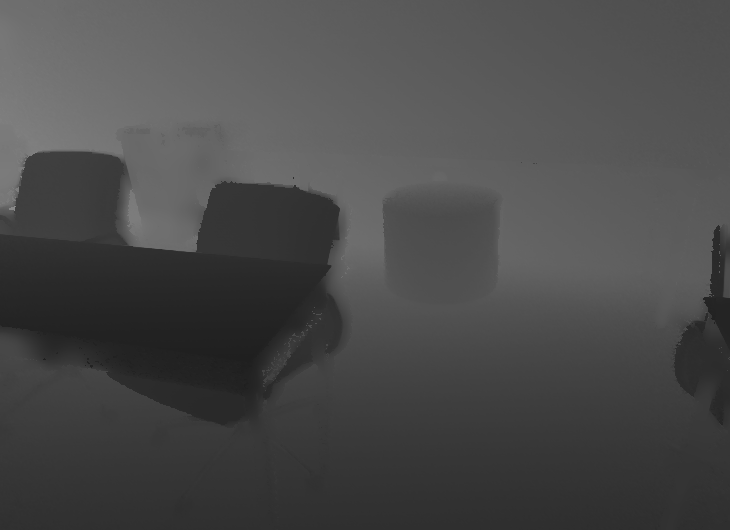}
   \end{minipage}
   \begin{minipage}[c]{0.091\linewidth}
      \includegraphics[width=0.66in]{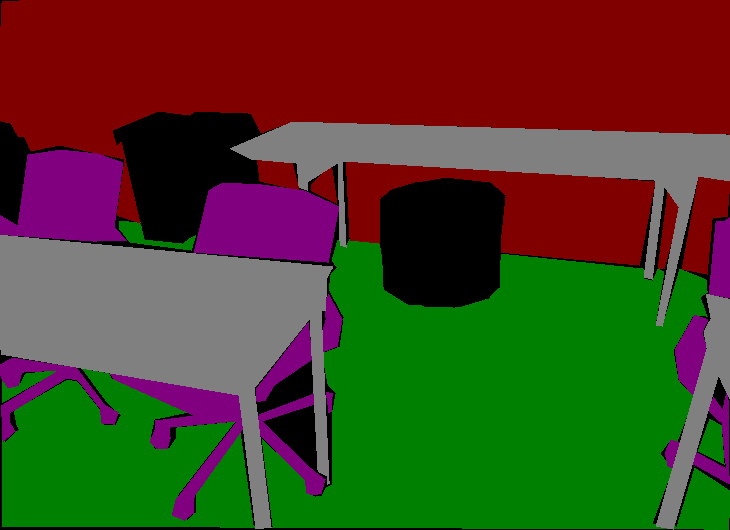}
   \end{minipage}
   \begin{minipage}[c]{0.091\linewidth}
      \includegraphics[width=0.66in]{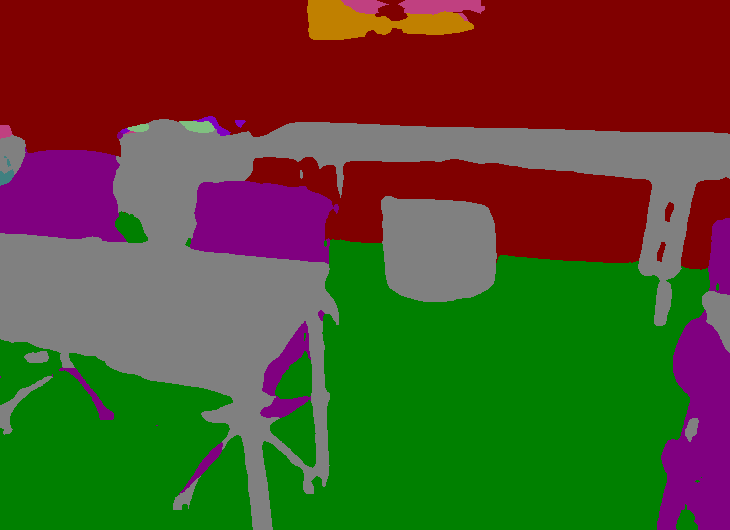}
   \end{minipage}
   \begin{minipage}[c]{0.091\linewidth}
      \includegraphics[width=0.66in]{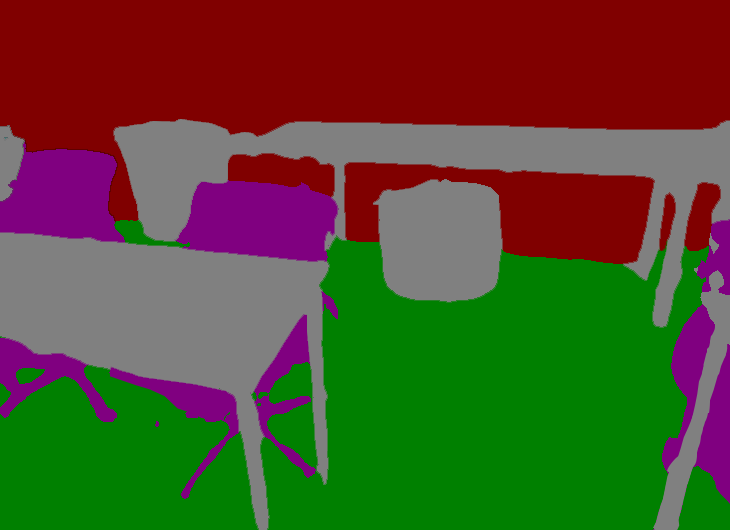}
   \end{minipage}

   %3
   \begin{minipage}[c]{0.091\linewidth}
      \includegraphics[width=0.66in]{pic/visualize/nyuv2/359_rgb.jpg}
   \end{minipage}
   \begin{minipage}[c]{0.091\linewidth}
      \includegraphics[width=0.66in]{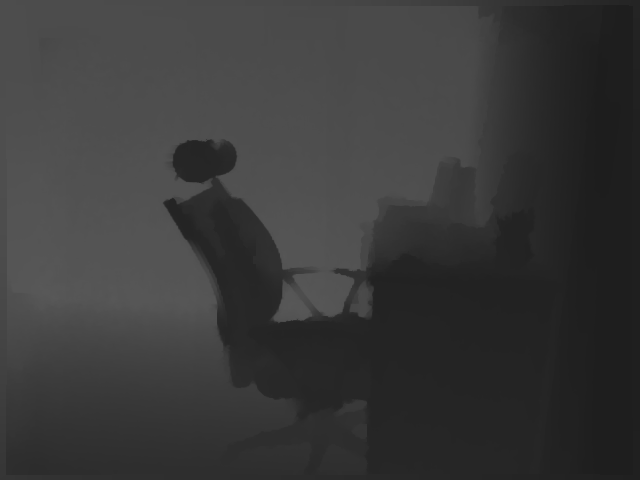}
   \end{minipage}
   \begin{minipage}[c]{0.091\linewidth}
      \includegraphics[width=0.66in]{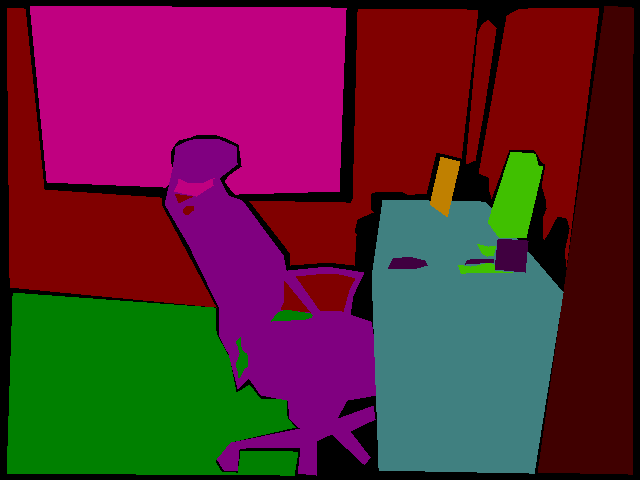}
   \end{minipage}
   \begin{minipage}[c]{0.091\linewidth}
      \includegraphics[width=0.66in]{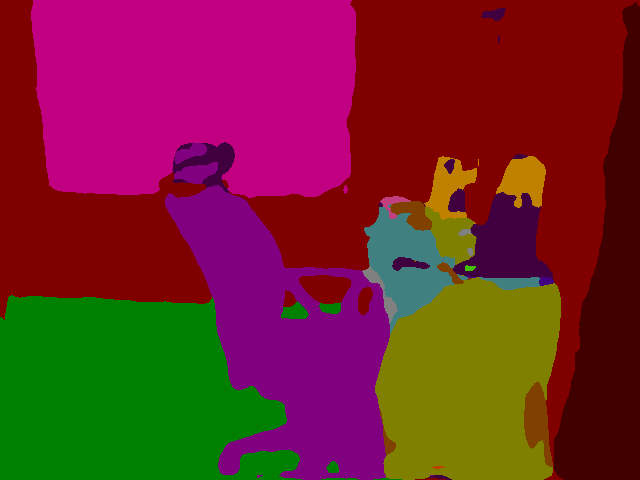}
   \end{minipage}
   \begin{minipage}[c]{0.091\linewidth}
      \includegraphics[width=0.66in]{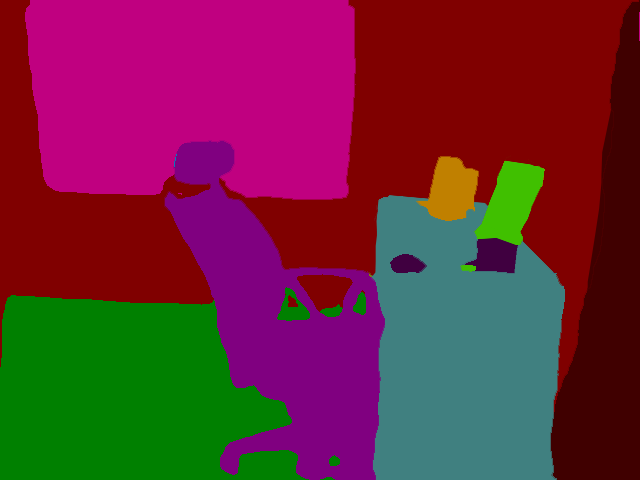}
   \end{minipage}\hspace{3pt}
   \begin{minipage}[c]{0.091\linewidth}
      \includegraphics[width=0.66in]{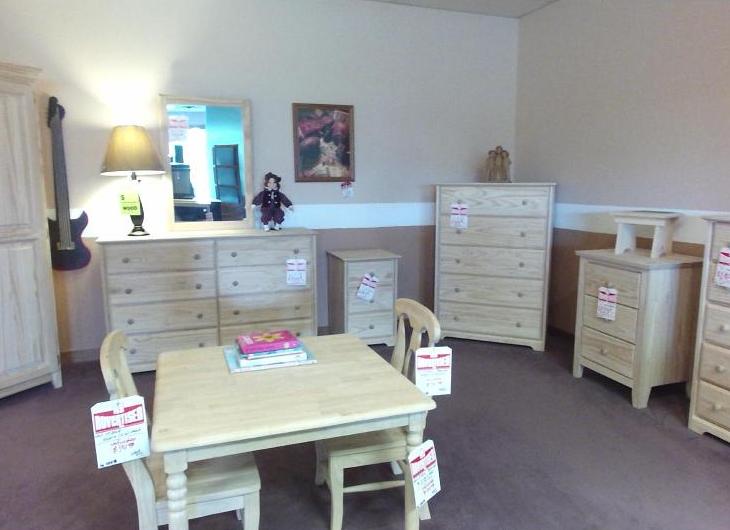}
   \end{minipage}
   \begin{minipage}[c]{0.091\linewidth}
      \includegraphics[width=0.66in]{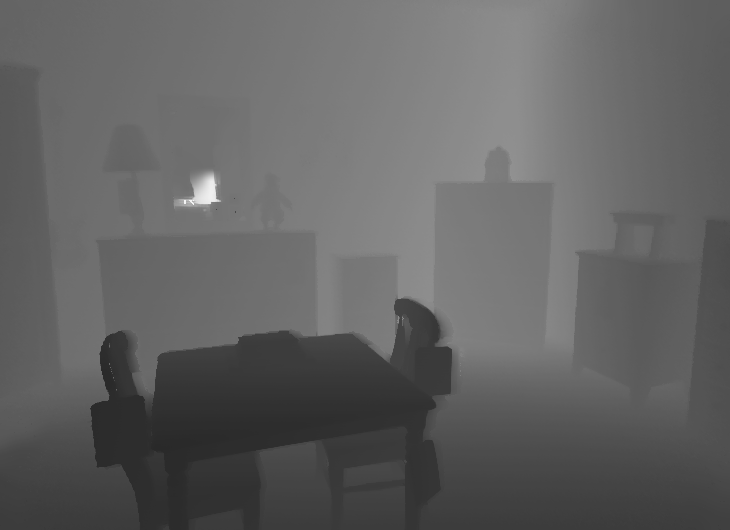}
   \end{minipage}
   \begin{minipage}[c]{0.091\linewidth}
      \includegraphics[width=0.66in]{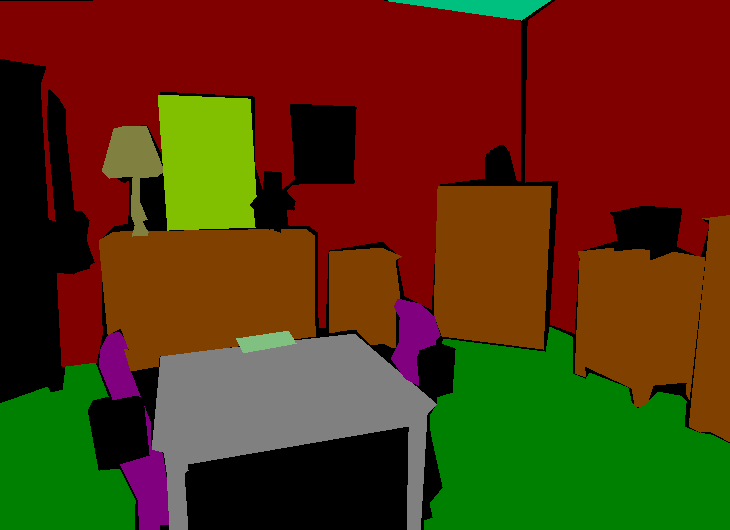}
   \end{minipage}
   \begin{minipage}[c]{0.091\linewidth}
      \includegraphics[width=0.66in]{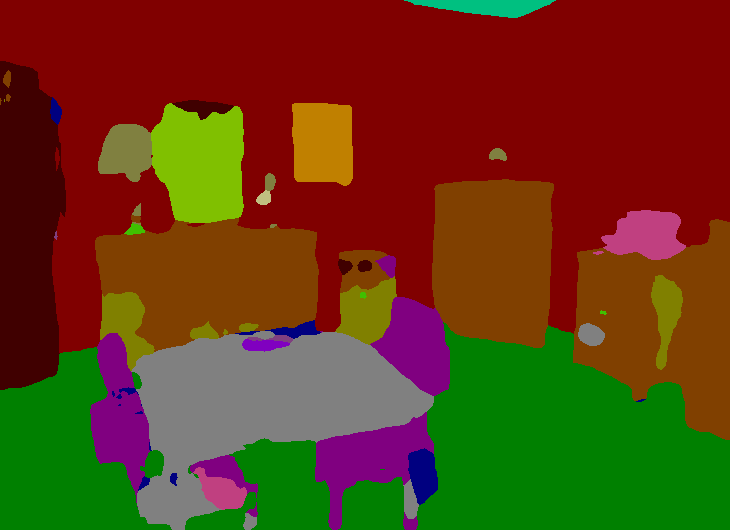}
   \end{minipage}
   \begin{minipage}[c]{0.091\linewidth}
      \includegraphics[width=0.66in]{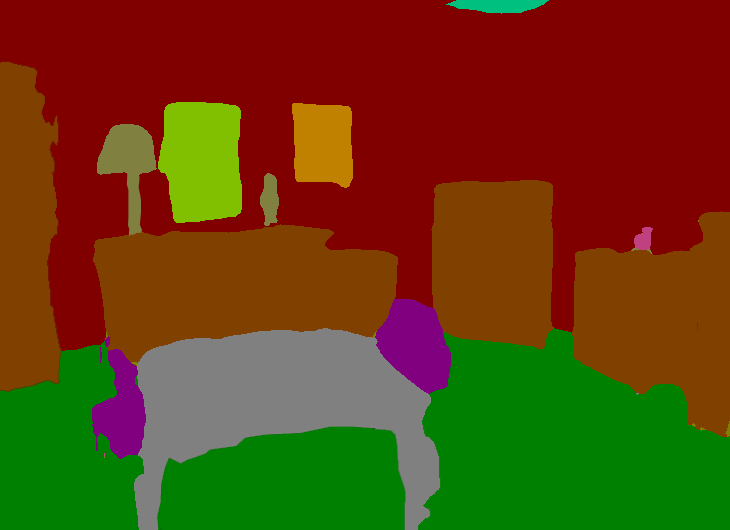}
   \end{minipage}

   %4
   \begin{minipage}[c]{0.091\linewidth}
      \includegraphics[width=0.66in]{pic/visualize/nyuv2/470_rgb.jpg}
      \centerline{ RGB }
   \end{minipage}
   \begin{minipage}[c]{0.091\linewidth}
      \includegraphics[width=0.66in]{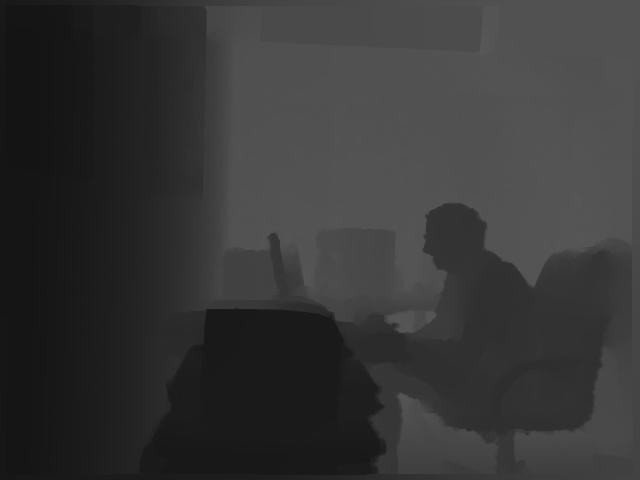}
      \centerline{ Depth }
   \end{minipage}
   \begin{minipage}[c]{0.091\linewidth}
      \includegraphics[width=0.66in]{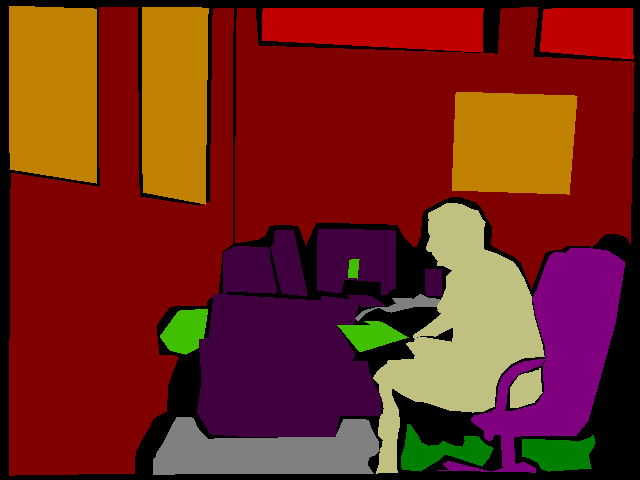}
      \centerline{GT}
   \end{minipage}
   \begin{minipage}[c]{0.091\linewidth}
      \includegraphics[width=0.66in]{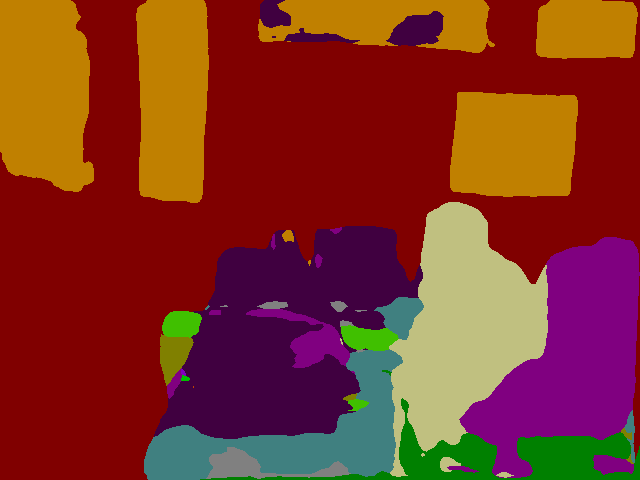}
      \centerline{Baseline }
   \end{minipage}
   \begin{minipage}[c]{0.091\linewidth}
      \includegraphics[width=0.66in]{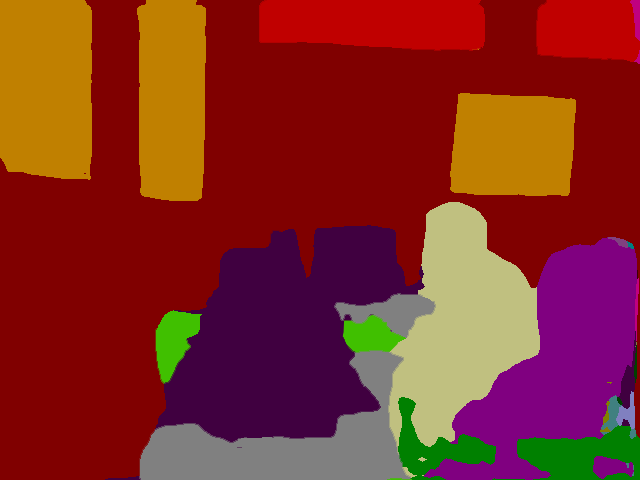}
      \centerline{Ours }
   \end{minipage}\hspace{3pt}
   \begin{minipage}[c]{0.091\linewidth}
      \includegraphics[width=0.66in]{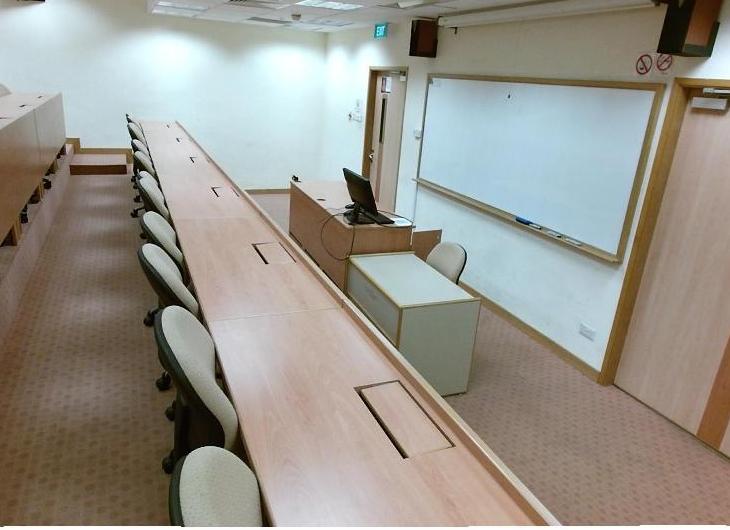}
      \centerline{RGB }
   \end{minipage}
   \begin{minipage}[c]{0.091\linewidth}
      \includegraphics[width=0.66in]{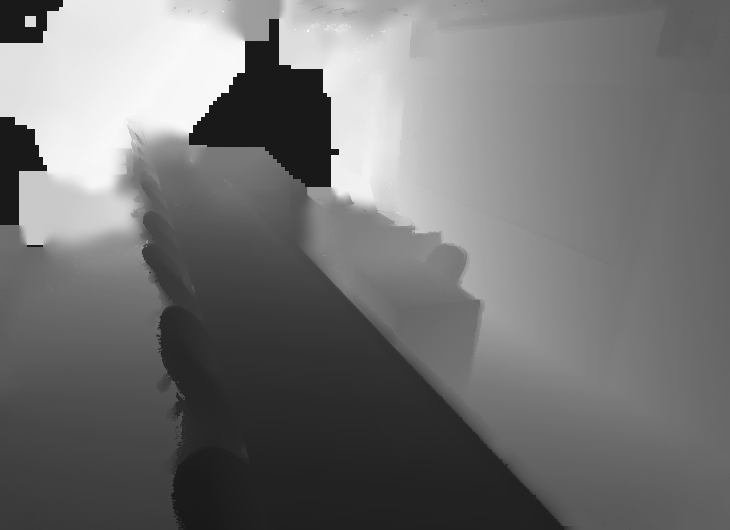}
      \centerline{ Depth }
   \end{minipage}
   \begin{minipage}[c]{0.091\linewidth}
      \includegraphics[width=0.66in]{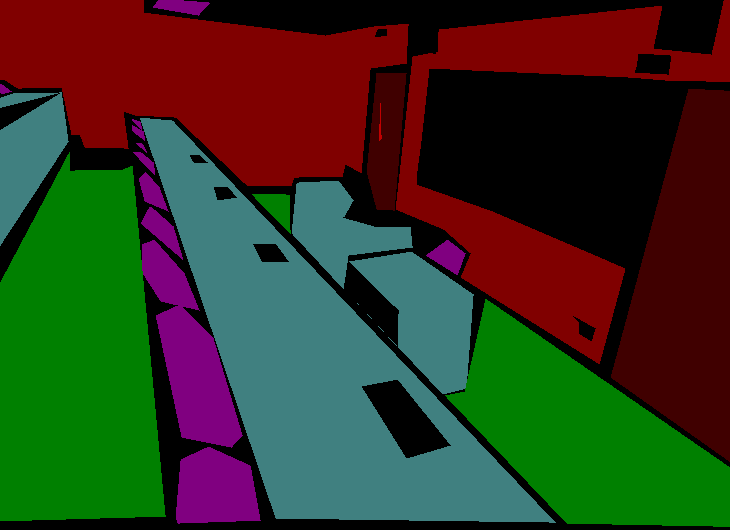}
      \centerline{ GT}
   \end{minipage}
   \begin{minipage}[c]{0.091\linewidth}
      \includegraphics[width=0.66in]{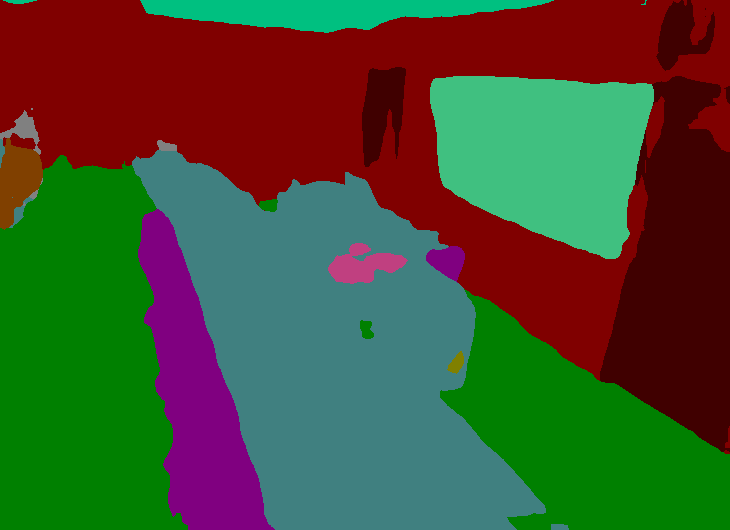}
      \centerline{Baseline }
   \end{minipage}
   \begin{minipage}[c]{0.091\linewidth}
      \includegraphics[width=0.66in]{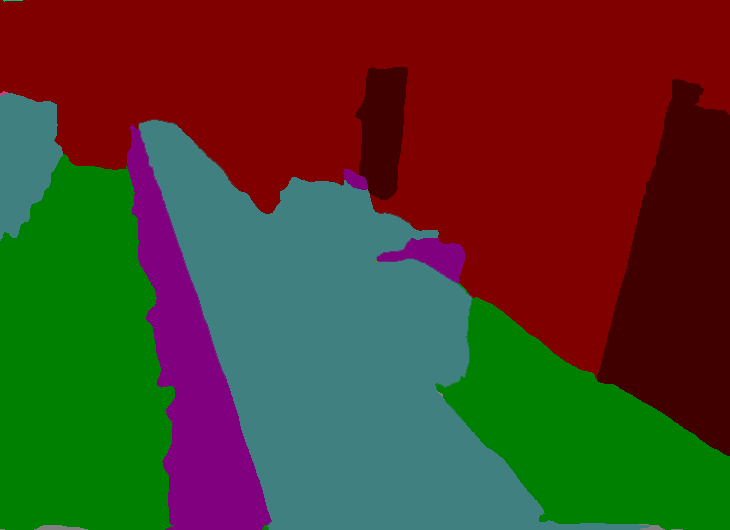}
      \centerline{Ours }
   \end{minipage}
   \caption{Visual comparison of scene semantic segmentation. The left part and right part are the testing results on the NYUDv2 dataset and SUN RGB-D dataset, respectively.}
   \label{visualization}
\end{figure*}

\begin{figure*}[t]
  \centering
   %1
   \begin{minipage}[c]{0.115\linewidth}
      \includegraphics[width=0.82in]{pic/visualize/350_feature_map/350_rgb.jpg}
   \end{minipage}
   \begin{minipage}[c]{0.115\linewidth}
      \includegraphics[width=0.82in]{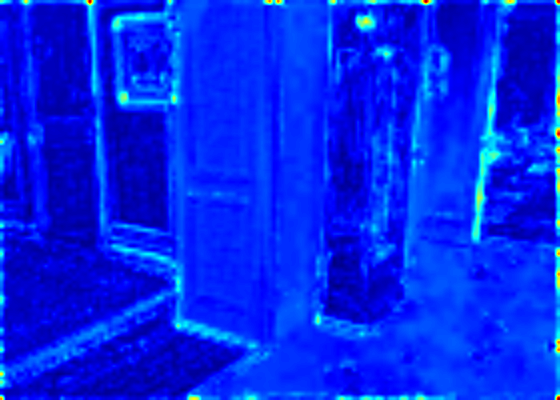}
   \end{minipage}
   \begin{minipage}[c]{0.115\linewidth}
      \includegraphics[width=0.82in]{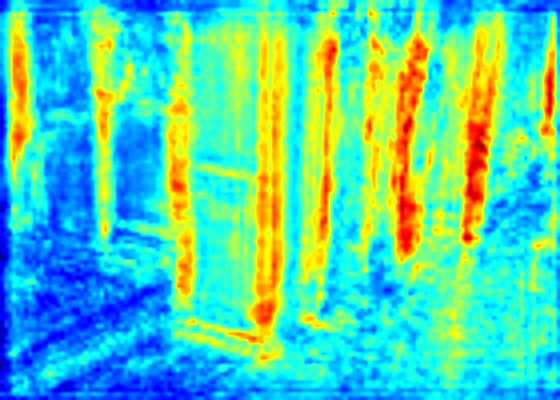}
   \end{minipage}
   \begin{minipage}[c]{0.115\linewidth}
      \includegraphics[width=0.82in]{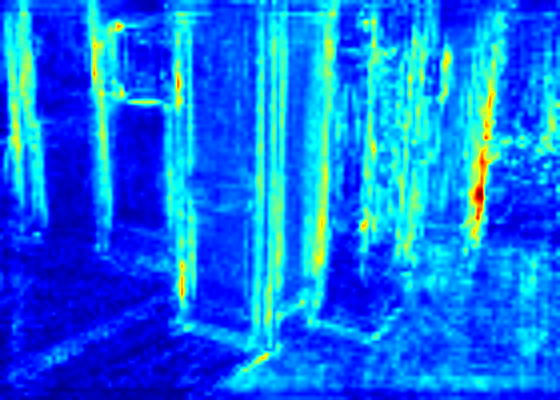}
   \end{minipage}\hspace{3pt}
      \begin{minipage}[c]{0.115\linewidth}
      \includegraphics[width=0.82in]{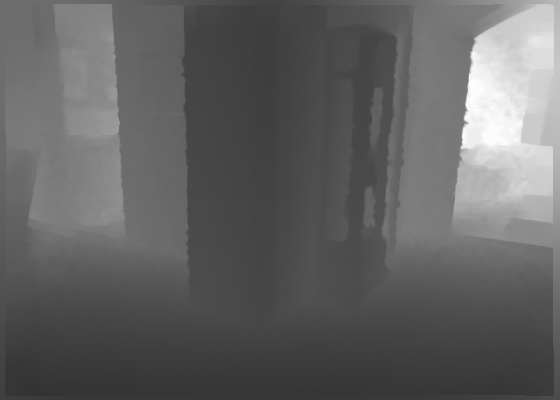}
   \end{minipage}
   \begin{minipage}[c]{0.115\linewidth}
      \includegraphics[width=0.82in]{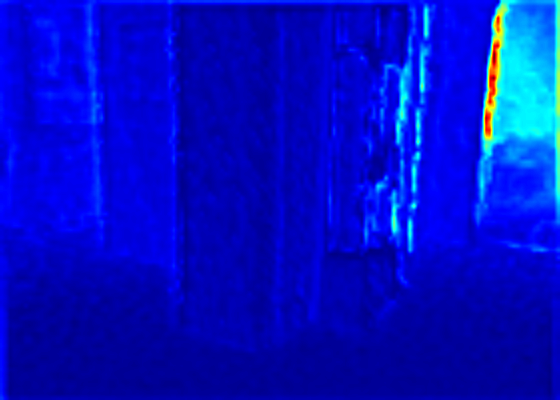}
   \end{minipage}
   \begin{minipage}[c]{0.115\linewidth}
      \includegraphics[width=0.82in]{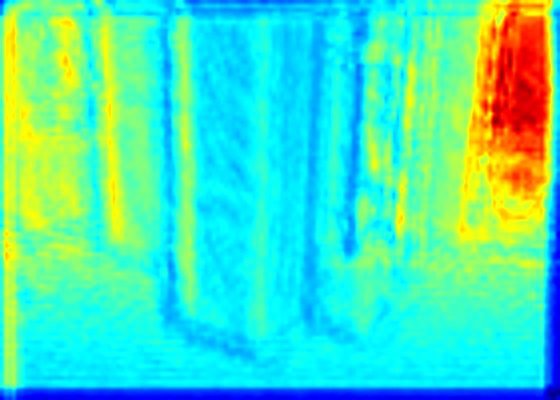}
   \end{minipage}
   \begin{minipage}[c]{0.115\linewidth}
      \includegraphics[width=0.82in]{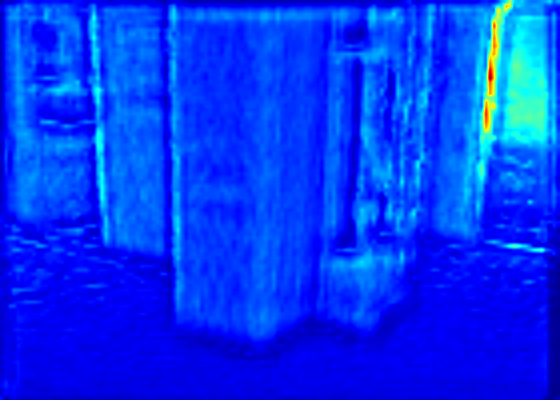}
   \end{minipage}

   \begin{minipage}[c]{0.115\linewidth}
      \includegraphics[width=0.82in]{pic/visualize/359_feature_map/359_rgb.jpg}
   \end{minipage}
   \begin{minipage}[c]{0.115\linewidth}
      \includegraphics[width=0.82in]{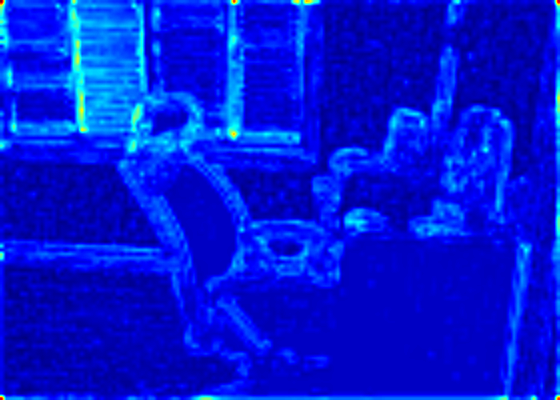}
   \end{minipage}
   \begin{minipage}[c]{0.115\linewidth}
      \includegraphics[width=0.82in]{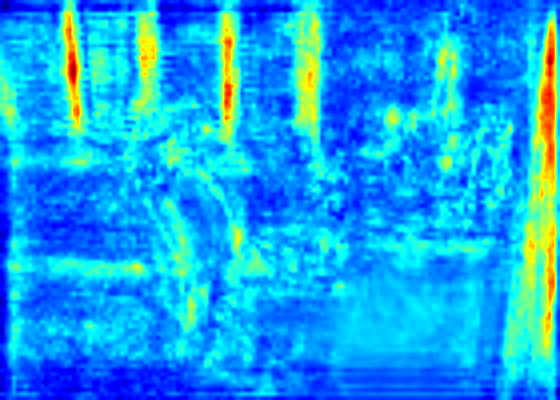}
   \end{minipage}
   \begin{minipage}[c]{0.115\linewidth}
      \includegraphics[width=0.82in]{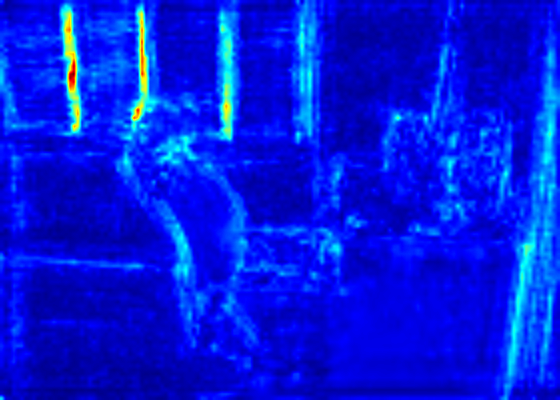}
   \end{minipage}\hspace{3pt}
      \begin{minipage}[c]{0.115\linewidth}
      \includegraphics[width=0.82in]{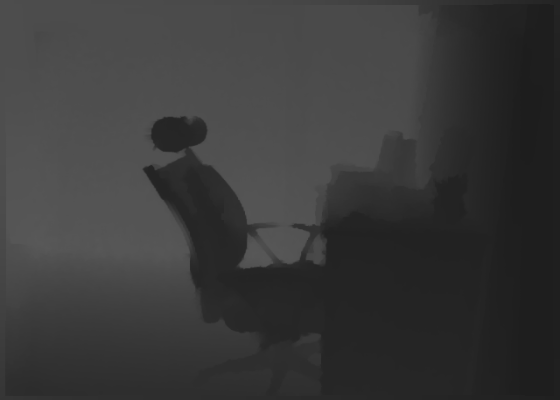}
   \end{minipage}
   \begin{minipage}[c]{0.115\linewidth}
      \includegraphics[width=0.82in]{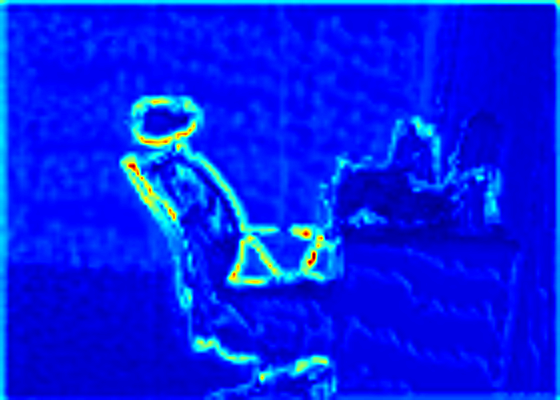}
   \end{minipage}
   \begin{minipage}[c]{0.115\linewidth}
      \includegraphics[width=0.82in]{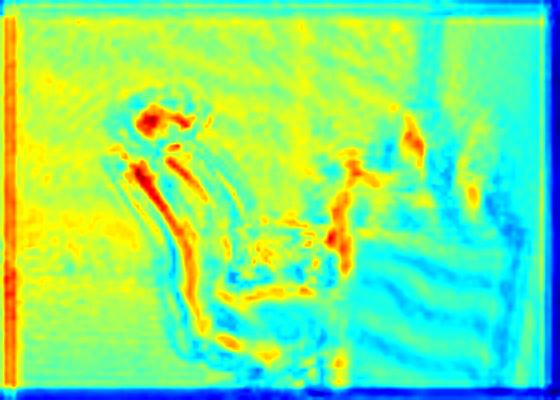}
   \end{minipage}
   \begin{minipage}[c]{0.115\linewidth}
      \includegraphics[width=0.82in]{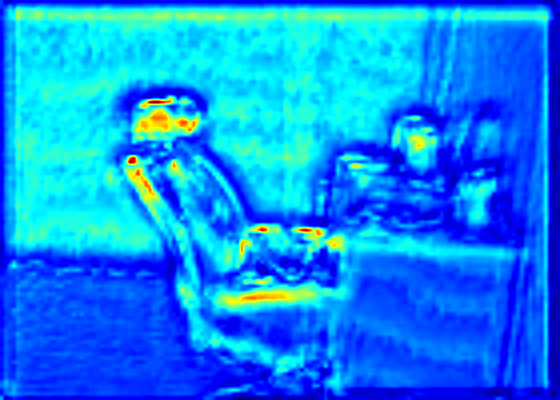}
   \end{minipage}

   \begin{minipage}[c]{0.115\linewidth}
      \includegraphics[width=0.82in]{pic/visualize/1423_feature_map/1423.jpg}
      \centerline{RGB}
   \end{minipage}
   \begin{minipage}[c]{0.115\linewidth}
      \includegraphics[width=0.82in]{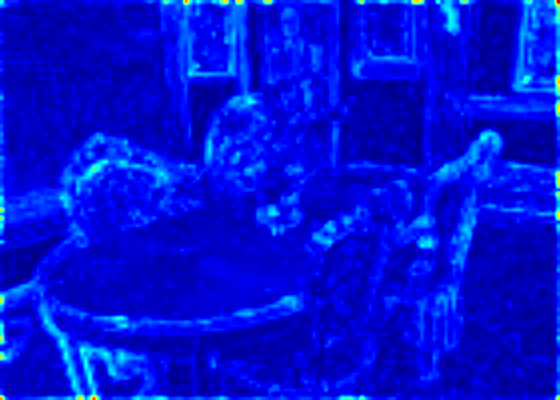}
      \centerline{W/O attention}
   \end{minipage}
   \begin{minipage}[c]{0.115\linewidth}
      \includegraphics[width=0.82in]{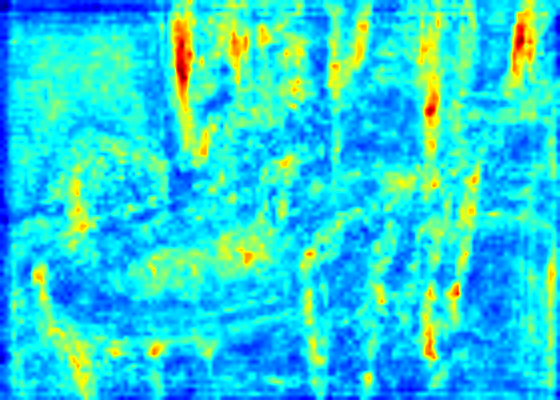}
      \centerline{Attention}
   \end{minipage}
   \begin{minipage}[c]{0.115\linewidth}
      \includegraphics[width=0.82in]{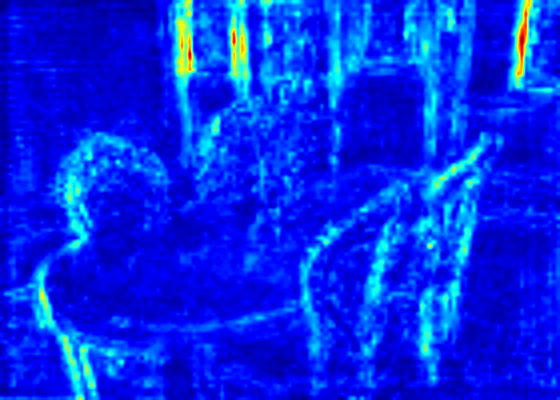}
      \centerline{Refined}
   \end{minipage}\hspace{3pt}
      \begin{minipage}[c]{0.115\linewidth}
      \includegraphics[width=0.82in]{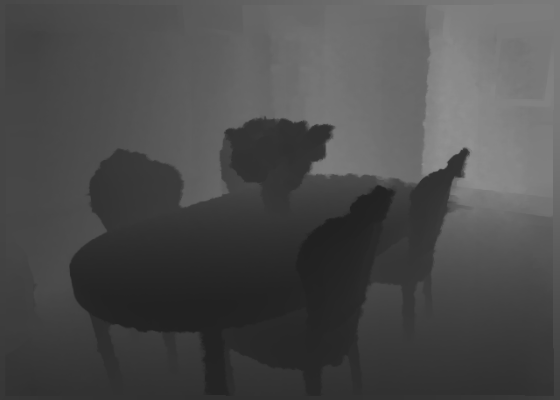}
      \centerline{Depth}
   \end{minipage}
   \begin{minipage}[c]{0.115\linewidth}
      \includegraphics[width=0.82in]{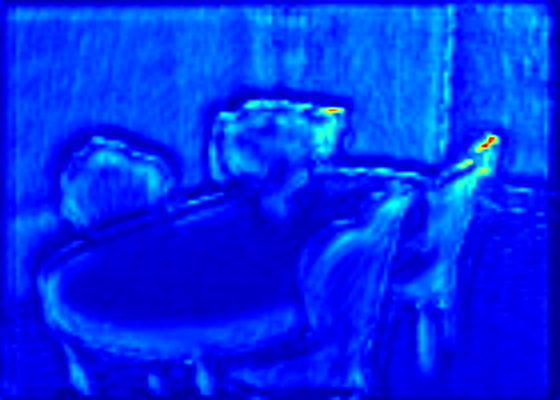}
      \centerline{W/O attention}
   \end{minipage}
   \begin{minipage}[c]{0.115\linewidth}
      \includegraphics[width=0.82in]{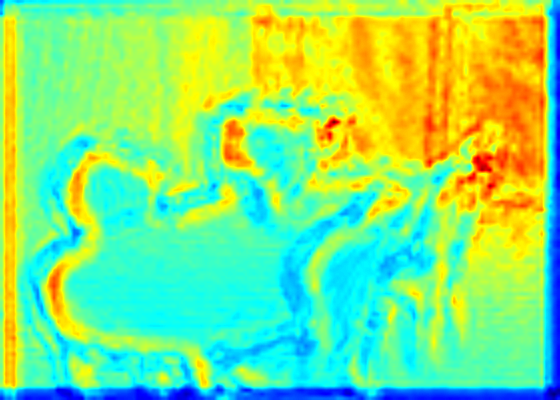}
      \centerline{Attention}
   \end{minipage}
   \begin{minipage}[c]{0.115\linewidth}
      \includegraphics[width=0.82in]{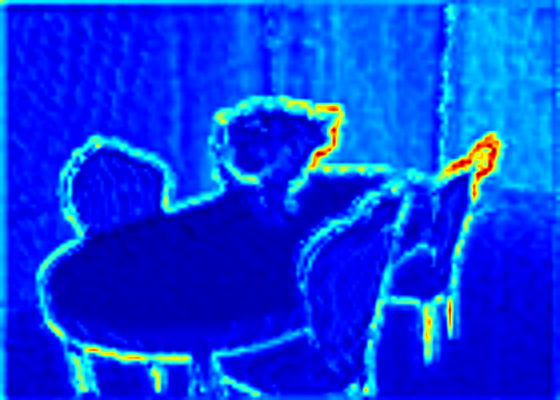}
      \centerline{Refined}
   \end{minipage}
   \caption{Visualization of response maps. For the left part, the second column shows the feature maps generated by baseline; the third column is generated by EDCA (the counterpart of the right part is generated by DCA) ; the fourth column shows the refined feature maps.}
   \label{featuremap}
\end{figure*}

\subsection{Comparing with State-of-the-arts}
% \subsubsection{NYUDv2}
\textbf{NYUDv2.} The comparison results are shown in Tab.~\ref{nyuv2}. Our method achieves leading performance. Compared with these methods,
our model focuses more on the variability within RGB and depth data and apply different modules to enhance the feature
representation. The depth-aware convolution proposed by D-CNN \cite{wang2018depth} is more similar to our approach. For fair comparison, under single testing, D-CNN achieves the mIoU of 48.4, while our model achieve the socre of 52.1, a 3.7\% improvement. This because
depth-aware convolution is used to produce a feature map while our DCA and EDCA are employed to generate an attention map that
indicates the importance of different points. Moreover, depth-aware convolution only compares the similarity of local regions
in the depth map and ignores the long-range dependence and global spatial consistency in RGB data, which can be captured by our EDCA.

% \subsubsection{SUN RGB-D}
\textbf{SUN RGB-D.}
Tab.~\ref{sunrgbd} shows testing results on SUN RGB-D dataset. The DCANet achieves the best results compared with other state-of-the-art methods under both single-scale and multi-scale testing.

\section{Visualization of DCANet}
Fig.~\ref{visualization} illustrates the qualitative results of the NYUDv2 and SUN RGB-D dataset. From the results, we can confirm
that the local geometric information in depth image and global dependence in RGB image are well enhanced by our DCA and EDCA modules. As illustrated in the second row on the right part, our DCANet successfully recognize the whole lamp including its bracket while it is even unrecognizable under strong lighting conditions. That is because our model effectively combines the advantages of both modal data. Specifically, when the 2D information of the object is unreliable, the model will make reasonable use of the corresponding geometric information. Similar examples can be found in
the second row on the left part.

To validate the effectiveness of the DCA and EDCA of our model, we apply the response maps of the baseline model and our DCANet. As shown in Fig.~\ref{featuremap}, the refined feature maps demonstrate the segmentation effectiveness of our method on capturing pixel-level subtle information (edge areas), where
the pixel differential convolution matters. The attention maps of the RGB and depth data also explain that the DCA provide intrinsic fine-grained local geometry difference information for depth data, while EDCA effectively provide a global view for RGB data.

\section{Conclusion}
Considering the intrinsic difference between RGB and depth data, we present a state-of-the-art differential convolution attention network by
introducing two plug-and-play modules: DCA and EDCA.
DCA dynamically perceives subtle geometric information that occurs in local regions in depth data. EDCA absorbs the advantage of dynamic convolution of DCA to propagate long-range
contextual dependencies and seamlessly incorporate spatial distribution for RGB data. Attention maps generated by DCA and EDCA are employed
to boost feature representations and further improve model performances.

%%%%%%%%% REFERENCES
{\small
\bibliographystyle{ieee_fullname}
% \bibliography{egbib}

\begin{thebibliography}{10}\itemsep=-1pt

\bibitem{badrinarayanan2017segnet}
Vijay Badrinarayanan, Alex Kendall, and Roberto Cipolla.
\newblock Segnet: A deep convolutional encoder-decoder architecture for image
  segmentation.
\newblock {\em IEEE transactions on pattern analysis and machine intelligence},
  39(12):2481--2495, 2017.

\bibitem{borse2021inverseform}
Shubhankar Borse, Ying Wang, Yizhe Zhang, and Fatih Porikli.
\newblock Inverseform: A loss function for structured boundary-aware
  segmentation.
\newblock In {\em Proceedings of the IEEE/CVF Conference on Computer Vision and
  Pattern Recognition}, pages 5901--5911, 2021.

\bibitem{cao2021shapeconv}
Jinming Cao, Hanchao Leng, Dani Lischinski, Daniel Cohen-Or, Changhe Tu, and
  Yangyan Li.
\newblock Shapeconv: Shape-aware convolutional layer for indoor rgb-d semantic
  segmentation.
\newblock In {\em Proceedings of the IEEE/CVF International Conference on
  Computer Vision}, pages 7088--7097, 2021.

\bibitem{cao2019gcnet}
Yue Cao, Jiarui Xu, Stephen Lin, Fangyun Wei, and Han Hu.
\newblock Gcnet: Non-local networks meet squeeze-excitation networks and
  beyond.
\newblock In {\em Proceedings of the IEEE/CVF International Conference on
  Computer Vision Workshops}, pages 0--0, 2019.

\bibitem{chen2017sca}
Long Chen, Hanwang Zhang, Jun Xiao, Liqiang Nie, Jian Shao, Wei Liu, and
  Tat-Seng Chua.
\newblock Sca-cnn: Spatial and channel-wise attention in convolutional networks
  for image captioning.
\newblock In {\em Proceedings of the IEEE conference on computer vision and
  pattern recognition}, pages 5659--5667, 2017.

\bibitem{chen2017deeplab}
Liang-Chieh Chen, George Papandreou, Iasonas Kokkinos, Kevin Murphy, and Alan~L
  Yuille.
\newblock Deeplab: Semantic image segmentation with deep convolutional nets,
  atrous convolution, and fully connected crfs.
\newblock {\em IEEE transactions on pattern analysis and machine intelligence},
  40(4):834--848, 2017.

\bibitem{chen2017rethinking}
Liang-Chieh Chen, George Papandreou, Florian Schroff, and Hartwig Adam.
\newblock Rethinking atrous convolution for semantic image segmentation.
\newblock {\em arXiv preprint arXiv:1706.05587}, 2017.

\bibitem{chen2018encoder}
Liang-Chieh Chen, Yukun Zhu, George Papandreou, Florian Schroff, and Hartwig
  Adam.
\newblock Encoder-decoder with atrous separable convolution for semantic image
  segmentation.
\newblock In {\em Proceedings of the European conference on computer vision
  (ECCV)}, pages 801--818, 2018.

\bibitem{chen2021spatial}
Lin-Zhuo Chen, Zheng Lin, Ziqin Wang, Yong-Liang Yang, and Ming-Ming Cheng.
\newblock Spatial information guided convolution for real-time rgbd semantic
  segmentation.
\newblock {\em IEEE Transactions on Image Processing}, 30:2313--2324, 2021.

\bibitem{chen2020bi}
Xiaokang Chen, Kwan-Yee Lin, Jingbo Wang, Wayne Wu, Chen Qian, Hongsheng Li,
  and Gang Zeng.
\newblock Bi-directional cross-modality feature propagation with
  separation-and-aggregation gate for rgb-d semantic segmentation.
\newblock In {\em European Conference on Computer Vision}, pages 561--577.
  Springer, 2020.

\bibitem{cheng2017locality}
Yanhua Cheng, Rui Cai, Zhiwei Li, Xin Zhao, and Kaiqi Huang.
\newblock Locality-sensitive deconvolution networks with gated fusion for rgb-d
  indoor semantic segmentation.
\newblock In {\em Proceedings of the IEEE conference on computer vision and
  pattern recognition}, pages 3029--3037, 2017.

\bibitem{chollet2017xception}
Fran{\c{c}}ois Chollet.
\newblock Xception: Deep learning with depthwise separable convolutions.
\newblock In {\em Proceedings of the IEEE conference on computer vision and
  pattern recognition}, pages 1251--1258, 2017.

\bibitem{ding2020semantic}
Henghui Ding, Xudong Jiang, Bing Shuai, Ai~Qun Liu, and Gang Wang.
\newblock Semantic segmentation with context encoding and multi-path decoding.
\newblock {\em IEEE Transactions on Image Processing}, 29:3520--3533, 2020.

\bibitem{fooladgar2019multi}
Fahimeh Fooladgar and Shohreh Kasaei.
\newblock Multi-modal attention-based fusion model for semantic segmentation of
  rgb-depth images.
\newblock {\em arXiv preprint arXiv:1912.11691}, 2019.

\bibitem{fu2019dual}
Jun Fu, Jing Liu, Haijie Tian, Yong Li, Yongjun Bao, Zhiwei Fang, and Hanqing
  Lu.
\newblock Dual attention network for scene segmentation.
\newblock In {\em Proceedings of the IEEE/CVF conference on computer vision and
  pattern recognition}, pages 3146--3154, 2019.

\bibitem{guo2022visual}
Meng-Hao Guo, Cheng-Ze Lu, Zheng-Ning Liu, Ming-Ming Cheng, and Shi-Min Hu.
\newblock Visual attention network.
\newblock {\em arXiv preprint arXiv:2202.09741}, 2022.

\bibitem{gupta2013perceptual}
Saurabh Gupta, Pablo Arbelaez, and Jitendra Malik.
\newblock Perceptual organization and recognition of indoor scenes from rgb-d
  images.
\newblock In {\em Proceedings of the IEEE conference on computer vision and
  pattern recognition}, pages 564--571, 2013.

\bibitem{gupta2014learning}
Saurabh Gupta, Ross Girshick, Pablo Arbel{\'a}ez, and Jitendra Malik.
\newblock Learning rich features from rgb-d images for object detection and
  segmentation.
\newblock In {\em European conference on computer vision}, pages 345--360.
  Springer, 2014.

\bibitem{he2016deep}
Kaiming He, Xiangyu Zhang, Shaoqing Ren, and Jian Sun.
\newblock Deep residual learning for image recognition.
\newblock In {\em Proceedings of the IEEE conference on computer vision and
  pattern recognition}, pages 770--778, 2016.

\bibitem{he2017std2p}
Yang He, Wei-Chen Chiu, Margret Keuper, and Mario Fritz.
\newblock Std2p: Rgbd semantic segmentation using spatio-temporal data-driven
  pooling.
\newblock In {\em Proceedings of the IEEE Conference on Computer Vision and
  Pattern Recognition}, pages 4837--4846, 2017.

\bibitem{hu2018gather}
Jie Hu, Li Shen, Samuel Albanie, Gang Sun, and Andrea Vedaldi.
\newblock Gather-excite: Exploiting feature context in convolutional neural
  networks.
\newblock {\em Advances in neural information processing systems}, 31, 2018.

\bibitem{hu2018squeeze}
Jie Hu, Li Shen, and Gang Sun.
\newblock Squeeze-and-excitation networks.
\newblock In {\em Proceedings of the IEEE conference on computer vision and
  pattern recognition}, pages 7132--7141, 2018.

\bibitem{hu2019acnet}
Xinxin Hu, Kailun Yang, Lei Fei, and Kaiwei Wang.
\newblock Acnet: Attention based network to exploit complementary features for
  rgbd semantic segmentation.
\newblock In {\em 2019 IEEE International Conference on Image Processing
  (ICIP)}, pages 1440--1444. IEEE, 2019.

\bibitem{husain2016combining}
Farzad Husain, Hannes Schulz, Babette Dellen, Carme Torras, and Sven Behnke.
\newblock Combining semantic and geometric features for object class
  segmentation of indoor scenes.
\newblock {\em IEEE Robotics and Automation Letters}, 2(1):49--55, 2016.

\bibitem{ioffe2015batch}
Sergey Ioffe and Christian Szegedy.
\newblock Batch normalization: Accelerating deep network training by reducing
  internal covariate shift.
\newblock In {\em International conference on machine learning}, pages
  448--456. PMLR, 2015.

\bibitem{jiang2018rednet}
Jindong Jiang, Lunan Zheng, Fei Luo, and Zhijun Zhang.
\newblock Rednet: Residual encoder-decoder network for indoor rgb-d semantic
  segmentation.
\newblock {\em arXiv preprint arXiv:1806.01054}, 2018.

\bibitem{jiao2019geometry}
Jianbo Jiao, Yunchao Wei, Zequn Jie, Honghui Shi, Rynson~WH Lau, and Thomas~S
  Huang.
\newblock Geometry-aware distillation for indoor semantic segmentation.
\newblock In {\em Proceedings of the IEEE/CVF Conference on Computer Vision and
  Pattern Recognition}, pages 2869--2878, 2019.

\bibitem{khan2016integrating}
Salman~H Khan, Mohammed Bennamoun, Ferdous Sohel, Roberto Togneri, and Imran
  Naseem.
\newblock Integrating geometrical context for semantic labeling of indoor
  scenes using rgbd images.
\newblock {\em International Journal of Computer Vision}, 117(1):1--20, 2016.

\bibitem{lin2017cascaded}
Di Lin, Guangyong Chen, Daniel Cohen-Or, Pheng-Ann Heng, and Hui Huang.
\newblock Cascaded feature network for semantic segmentation of rgb-d images.
\newblock In {\em Proceedings of the IEEE international conference on computer
  vision}, pages 1311--1319, 2017.

\bibitem{lin2017refinenet}
Guosheng Lin, Anton Milan, Chunhua Shen, and Ian Reid.
\newblock Refinenet: Multi-path refinement networks for high-resolution
  semantic segmentation.
\newblock In {\em Proceedings of the IEEE conference on computer vision and
  pattern recognition}, pages 1925--1934, 2017.

\bibitem{lin2016efficient}
Guosheng Lin, Chunhua Shen, Anton Van Den~Hengel, and Ian Reid.
\newblock Efficient piecewise training of deep structured models for semantic
  segmentation.
\newblock In {\em Proceedings of the IEEE conference on computer vision and
  pattern recognition}, pages 3194--3203, 2016.

\bibitem{lin2013network}
Min Lin, Qiang Chen, and Shuicheng Yan.
\newblock Network in network.
\newblock {\em arXiv preprint arXiv:1312.4400}, 2013.

\bibitem{lin2017feature}
Tsung-Yi Lin, Piotr Doll{\'a}r, Ross Girshick, Kaiming He, Bharath Hariharan,
  and Serge Belongie.
\newblock Feature pyramid networks for object detection.
\newblock In {\em Proceedings of the IEEE conference on computer vision and
  pattern recognition}, pages 2117--2125, 2017.

\bibitem{liu2015parsenet}
Wei Liu, Andrew Rabinovich, and Alexander~C Berg.
\newblock Parsenet: Looking wider to see better.
\newblock {\em arXiv preprint arXiv:1506.04579}, 2015.

\bibitem{long2015fully}
Jonathan Long, Evan Shelhamer, and Trevor Darrell.
\newblock Fully convolutional networks for semantic segmentation.
\newblock In {\em Proceedings of the IEEE conference on computer vision and
  pattern recognition}, pages 3431--3440, 2015.

\bibitem{mei2021depth}
Haiyang Mei, Bo Dong, Wen Dong, Pieter Peers, Xin Yang, Qiang Zhang, and
  Xiaopeng Wei.
\newblock Depth-aware mirror segmentation.
\newblock In {\em Proceedings of the IEEE/CVF Conference on Computer Vision and
  Pattern Recognition}, pages 3044--3053, 2021.

\bibitem{nair2010rectified}
Vinod Nair and Geoffrey~E Hinton.
\newblock Rectified linear units improve restricted boltzmann machines.
\newblock In {\em Icml}, 2010.

\bibitem{ojala2002multiresolution}
Timo Ojala, Matti Pietikainen, and Topi Maenpaa.
\newblock Multiresolution gray-scale and rotation invariant texture
  classification with local binary patterns.
\newblock {\em IEEE Transactions on pattern analysis and machine intelligence},
  24(7):971--987, 2002.

\bibitem{park2017rdfnet}
Seong-Jin Park, Ki-Sang Hong, and Seungyong Lee.
\newblock Rdfnet: Rgb-d multi-level residual feature fusion for indoor semantic
  segmentation.
\newblock In {\em Proceedings of the IEEE international conference on computer
  vision}, pages 4980--4989, 2017.

\bibitem{paszke2019pytorch}
Adam Paszke, Sam Gross, Francisco Massa, Adam Lerer, James Bradbury, Gregory
  Chanan, Trevor Killeen, Zeming Lin, Natalia Gimelshein, Luca Antiga, et~al.
\newblock Pytorch: An imperative style, high-performance deep learning library.
\newblock {\em Advances in neural information processing systems}, 32, 2019.

\bibitem{qi2019amodal}
Lu Qi, Li Jiang, Shu Liu, Xiaoyong Shen, and Jiaya Jia.
\newblock Amodal instance segmentation with kins dataset.
\newblock In {\em Proceedings of the IEEE/CVF Conference on Computer Vision and
  Pattern Recognition}, pages 3014--3023, 2019.

\bibitem{qi20173d}
Xiaojuan Qi, Renjie Liao, Jiaya Jia, Sanja Fidler, and Raquel Urtasun.
\newblock 3d graph neural networks for rgbd semantic segmentation.
\newblock In {\em Proceedings of the IEEE International Conference on Computer
  Vision}, pages 5199--5208, 2017.

\bibitem{qin2020ffa}
Xu Qin, Zhilin Wang, Yuanchao Bai, Xiaodong Xie, and Huizhu Jia.
\newblock Ffa-net: Feature fusion attention network for single image dehazing.
\newblock In {\em Proceedings of the AAAI Conference on Artificial
  Intelligence}, volume~34, pages 11908--11915, 2020.

\bibitem{ramachandran2019stand}
Prajit Ramachandran, Niki Parmar, Ashish Vaswani, Irwan Bello, Anselm Levskaya,
  and Jon Shlens.
\newblock Stand-alone self-attention in vision models.
\newblock {\em Advances in Neural Information Processing Systems}, 32, 2019.

\bibitem{ren2012rgb}
Xiaofeng Ren, Liefeng Bo, and Dieter Fox.
\newblock Rgb-(d) scene labeling: Features and algorithms.
\newblock In {\em 2012 IEEE Conference on Computer Vision and Pattern
  Recognition}, pages 2759--2766. IEEE, 2012.

\bibitem{russakovsky2015imagenet}
Olga Russakovsky, Jia Deng, Hao Su, Jonathan Krause, Sanjeev Satheesh, Sean Ma,
  Zhiheng Huang, Andrej Karpathy, Aditya Khosla, Michael Bernstein, et~al.
\newblock Imagenet large scale visual recognition challenge.
\newblock {\em International journal of computer vision}, 115(3):211--252,
  2015.

\bibitem{silberman2012indoor}
Nathan Silberman, Derek Hoiem, Pushmeet Kohli, and Rob Fergus.
\newblock Indoor segmentation and support inference from rgbd images.
\newblock In {\em European conference on computer vision}, pages 746--760.
  Springer, 2012.

\bibitem{song2015sun}
Shuran Song, Samuel~P Lichtenberg, and Jianxiong Xiao.
\newblock Sun rgb-d: A rgb-d scene understanding benchmark suite.
\newblock In {\em Proceedings of the IEEE conference on computer vision and
  pattern recognition}, pages 567--576, 2015.

\bibitem{srivastava2014dropout}
Nitish Srivastava, Geoffrey Hinton, Alex Krizhevsky, Ilya Sutskever, and Ruslan
  Salakhutdinov.
\newblock Dropout: a simple way to prevent neural networks from overfitting.
\newblock {\em The journal of machine learning research}, 15(1):1929--1958,
  2014.

\bibitem{vaswani2021scaling}
Ashish Vaswani, Prajit Ramachandran, Aravind Srinivas, Niki Parmar, Blake
  Hechtman, and Jonathon Shlens.
\newblock Scaling local self-attention for parameter efficient visual
  backbones.
\newblock In {\em Proceedings of the IEEE/CVF Conference on Computer Vision and
  Pattern Recognition}, pages 12894--12904, 2021.

\bibitem{vaswani2017attention}
Ashish Vaswani, Noam Shazeer, Niki Parmar, Jakob Uszkoreit, Llion Jones,
  Aidan~N Gomez, {\L}ukasz Kaiser, and Illia Polosukhin.
\newblock Attention is all you need.
\newblock {\em Advances in neural information processing systems}, 30, 2017.

\bibitem{wang2017residual}
Fei Wang, Mengqing Jiang, Chen Qian, Shuo Yang, Cheng Li, Honggang Zhang,
  Xiaogang Wang, and Xiaoou Tang.
\newblock Residual attention network for image classification.
\newblock In {\em Proceedings of the IEEE conference on computer vision and
  pattern recognition}, pages 3156--3164, 2017.

\bibitem{wang2018depth}
Weiyue Wang and Ulrich Neumann.
\newblock Depth-aware cnn for rgb-d segmentation.
\newblock In {\em Proceedings of the European Conference on Computer Vision
  (ECCV)}, pages 135--150, 2018.

\bibitem{wang2018non}
Xiaolong Wang, Ross Girshick, Abhinav Gupta, and Kaiming He.
\newblock Non-local neural networks.
\newblock In {\em Proceedings of the IEEE conference on computer vision and
  pattern recognition}, pages 7794--7803, 2018.

\bibitem{wang2022multimodal}
Yikai Wang, Xinghao Chen, Lele Cao, Wenbing Huang, Fuchun Sun, and Yunhe Wang.
\newblock Multimodal token fusion for vision transformers.
\newblock In {\em Proceedings of the IEEE/CVF Conference on Computer Vision and
  Pattern Recognition}, pages 12186--12195, 2022.

\bibitem{woo2018cbam}
Sanghyun Woo, Jongchan Park, Joon-Young Lee, and In~So Kweon.
\newblock Cbam: Convolutional block attention module.
\newblock In {\em Proceedings of the European conference on computer vision
  (ECCV)}, pages 3--19, 2018.

\bibitem{xie2021segformer}
Enze Xie, Wenhai Wang, Zhiding Yu, Anima Anandkumar, Jose~M Alvarez, and Ping
  Luo.
\newblock Segformer: Simple and efficient design for semantic segmentation with
  transformers.
\newblock {\em Advances in Neural Information Processing Systems}, 34, 2021.

\bibitem{xing2020malleable}
Yajie Xing, Jingbo Wang, and Gang Zeng.
\newblock Malleable 2.5 d convolution: Learning receptive fields along the
  depth-axis for rgb-d scene parsing.
\newblock In {\em European Conference on Computer Vision}, pages 555--571.
  Springer, 2020.

\bibitem{ye2022inverted}
Hanrong Ye and Dan Xu.
\newblock Inverted pyramid multi-task transformer for dense scene
  understanding.
\newblock {\em arXiv preprint arXiv:2203.07997}, 2022.

\bibitem{zhang2019self}
Han Zhang, Ian Goodfellow, Dimitris Metaxas, and Augustus Odena.
\newblock Self-attention generative adversarial networks.
\newblock In {\em International conference on machine learning}, pages
  7354--7363. PMLR, 2019.

\bibitem{zhao2020exploring}
Hengshuang Zhao, Jiaya Jia, and Vladlen Koltun.
\newblock Exploring self-attention for image recognition.
\newblock In {\em Proceedings of the IEEE/CVF Conference on Computer Vision and
  Pattern Recognition}, pages 10076--10085, 2020.

\bibitem{zhao2017pyramid}
Hengshuang Zhao, Jianping Shi, Xiaojuan Qi, Xiaogang Wang, and Jiaya Jia.
\newblock Pyramid scene parsing network.
\newblock In {\em Proceedings of the IEEE conference on computer vision and
  pattern recognition}, pages 2881--2890, 2017.

\bibitem{zhu2019asymmetric}
Zhen Zhu, Mengde Xu, Song Bai, Tengteng Huang, and Xiang Bai.
\newblock Asymmetric non-local neural networks for semantic segmentation.
\newblock In {\em Proceedings of the IEEE/CVF International Conference on
  Computer Vision}, pages 593--602, 2019.

\end{thebibliography}
}

\end{document}